\documentclass[notitlepage,
 aip,
 amsmath,amssymb,
reprint,%
]{revtex4-2}

\usepackage[version=3]{mhchem} 
\usepackage{multirow}
\usepackage{threeparttable}
\usepackage{appendix}
\usepackage{braket}
\usepackage{longtable}
\usepackage{color}
\usepackage{graphicx}
\usepackage{bm}
\usepackage[section]{placeins}
\usepackage{soul}
\usepackage{siunitx}
\usepackage[table]{xcolor}
\usepackage{amsmath}

\renewcommand{\thetable}{\arabic{table}}
\usepackage{multirow}
\usepackage{threeparttable}
\usepackage{eucal}

\begin{document}

\preprint{AIP/123-QED}

\title[]{
Improving Aufbau Suppressed Coupled Cluster
Through Perturbative Analysis
}

\author{Harrison Tuckman}
\affiliation{
Department of Chemistry, University of California, Berkeley, California 94720, USA 
}

\author{Ziheng Ma}
\affiliation{
Department of Chemistry, University of California, Berkeley, California 94720, USA 
}

\author{Eric Neuscamman}
\email{eneuscamman@berkeley.edu}
\affiliation{
Department of Chemistry, University of California, Berkeley, California 94720, USA 
}
\affiliation{Chemical Sciences Division, Lawrence Berkeley National Laboratory, Berkeley, CA, 94720, USA}

\date{\today}

\begin{abstract}
  
Guided by perturbative analysis, we improve the accuracy of
Aufbau suppressed coupled cluster theory
in simple single excitations, multi-configurational single excitations,
and charge transfer excitations while keeping the cost of its
leading-order terms precisely in line with ground state coupled cluster.
Combining these accuracy improvements with a more
efficient implementation based on spin-adaptation,
we observe high accuracy in a large test set of
single excitations, and, in particular, a mean unsigned error
for charge transfer states that outperforms
equation-of-motion coupled cluster theory by 0.25 eV.
We discuss how these results are achieved via a systematic identification
of which amplitudes to prioritize for single- and multi-configurational
excited states, and how this prioritization differs in important
ways from the ground state theory.
In particular, our data show that a partial linearization of the
theory increases accuracy by mitigating unwanted
side effects of Aufbau suppression.

\end{abstract}

\maketitle


\section{Introduction}

It is hard to overstate the crucial role that perturbative analysis has
played in making coupled cluster (CC) theory the dominant approach to
high accuracy modeling of weakly correlated molecular ground states.
\cite{bartlett2007coupled,crawford2007introduction,shavitt2009many}
When choosing which amplitudes to include at a given level of
theory or how to devise lower-cost ways to approximate the effects of
higher-body amplitudes, identifying amplitudes' and energy terms'
orders within the framework of many-body perturbation theory (MBPT)
is indispensable.
In this study, we begin the process of bringing this powerful analysis
to bear on Aufbau suppressed coupled cluster (ASCC), \cite{tuckman2024aufbau}
a recently introduced approach to excited-state-specific CC theory
whose costs and state-specificity closely mirror those of the
traditional ground state theory.
This similitude leads to many familiar parallels in the analysis, 
but there are also interesting differences that intersect with
our central design choice of ensuring that the $N^6$-scaling terms
in the theory retain strict cost parity with those
of ground state CC with singles and doubles (CCSD).
This study explores this intersection and how, through some
key modifications to ASCC, the analysis leads to substantial
accuracy improvements and, in particular, a clear accuracy
advantage in charge transfer states when compared to existing,
similar-cost excited state CC methods.

Recent years have seen considerable efforts in the area of excited-state-specific
electronic structure theory, including efforts in
single determinant methods,
\cite{ziegler1977calculation,gilbert2008self,besley2009self,barca2018simple,carter2020state,burton2020energy} 
configuration interaction (CI),
\cite{dreuw2005single,liu2012communication,liu2014variationally,shea2018communication,shea2020generalized,hardikar2020self,kossoski2022state,kossoski2023seniority,burton2022energy,tsuchimochi2024CISthenCIS}
perturbation theory, \cite{clune2020n5,clune2023studying} 
CC theory,
\cite{mayhall2010multiple,lee2019excited,kossoski2021excited,marie2021variational,rishi2023dark,damour2024state,tuckman2023excited,tuckman2024aufbau}
multi-reference theory,
\cite{knowles1985efficient,werner1985second,ruedenberg1982atoms,roos1987complete,tran2019tracking,tran2020improving,hanscam2022applying,saade2024excited}
and density functional theory (DFT).
\cite{kowalczyk2011assessment,kowalczyk2013excitation,hait2020excited,hait2021orbital,zhao2019density,levi2020variational,hait2020highly,kempfer2022role}
Compared to linear response theories like time-dependent DFT (TD-DFT)
\cite{runge1984density,burke2005time,casida2012progress}
and linear response (LR) and equation-of-motion (EOM) CC,
\cite{monkhorst1977calculation,dalgaard1983some,sekino1984linear,koch1990excitation,koch1990coupled,rico1993single,koch1994calculation,sneskov2012excited, rowe1968equations,stanton1993equation,krylov2008equation}
these state-specific approaches are typically better able to
incorporate post-excitation orbital relaxation effects and,
at least in principle, can achieve better balance between ground and excited
states by offering a fully tailored description of each state.
Of course, state-specific approaches must typically perform dedicated
nonlinear optimizations for each state,
while linear response theories typically allow many states
to be treated via a single linear diagonalization.
In many applications, this reality makes linear response preferable,
but state-specific approaches can offer substantially
improved accuracy, especially in cases like charge transfer and
core excitations where orbital relaxation effects are more significant. 
\cite{hait2020highly,tuckman2024aufbau,kozma2020new,subotnik2011communication}
With many such states existing as otherwise straightforward 
single excitations dominated by one or at most a few
configuration state functions (CSFs),
one wonders how closely a CC theory based on
an orbital-relaxed single- or few-CSF reference state
could mirror single-reference ground state CC,
and whether such an approach would be similarly effective
at capturing weak correlation.

ASCC attempts to answer this question by incorporating a de-excitation
exponential that, in concert with the traditional excitation exponential,
converts a closed-shell single-determinant formal reference state into
an expansion in which that determinant is suppressed or even absent.
Remarkably simple choices for both the excitation and de-excitation
operators produce single-CSF singly-excited states, at which point
the inclusion of amplitudes up to doubly excited relative to the
single excitation offers good accuracy for the weak correlation treatment,
in close analogy to CCSD. \cite{tuckman2024aufbau}
Preliminary results for multi-CSF states, however, were far less
encouraging, with eV-sized errors similar to those seen in the closely
related two-determinant CC approach.\cite{damour2024state}
This difficulty prompts the question of how to go about improving the theory.
Recognizing that the basic framework of ASCC is straightforwardly
systematically improvable --- as in ground state CC, adding higher and
higher excitations eventually recreates full configuration interaction (FCI) --- 
we turn in the present study to a perturbative analysis in order to
determine whether key terms or amplitudes can be addressed without
requiring any $N^6$ terms beyond those already present in CCSD.

We approach ASCC's perturbative analysis in much the same way as in
ground state theory and find many parallels, but the zeroth order
components of ASCC's cluster operators create differences that,
once identified, aid us in making three key improvements to the theory.
First, while ASCC's higher-body amplitudes mostly follow the
ground state pattern in which each additional excitation level increases
the perturbative order by one, key subsets of higher-body amplitudes
start out at a lower order than in the ground state.
As a result, a single-CSF state in ASCC requires a subset of the
quadruples if its energy is to be complete through third order.
These quadruples would lead to additional $N^6$ terms,
putting the goal of matching CCSD's third order
completeness in conflict with our design choice of avoiding
such terms.
We therefore adopt, for this study at least, a partial linearization
scheme to mitigate the effects of neglecting these quadruples.
Second, for multi-CSF states, we find that our original formulation
of ASCC was missing some first order amplitudes.
Happily, including them dramatically improves accuracy without adding
any additional $N^6$ terms.
Third, while carrying out this analysis, we also made the tangential
discovery that the basic ASCC formalism contains two subtly different
ansatz definitions.
For now, we adopt a practical averaging approach that avoids having
to choose between them.
In future work, it will be interesting to analyze this duality
further and explore additional MBPT-inspired improvements, such as
non-iterative approaches to third order energies and (T) analogues.
In the present study, however, we keep our focus on
improvements that can be made without any additional $N^6$ terms.

After this formal analysis, we leverage our newly spin-adapted implementation
to perform tests on a greatly expanded set of molecular excited states.
We begin with the excellent QUEST benchmarks of small- and medium-sized molecules'
valence \textcolor{black}{and Rydberg} excited states, \cite{loos2018mountaineering,loos2020mountaineering}
where we find that the improved ASCC matches EOM-CCSD in accuracy for
single-CSF states.
For two-CSF states, the improvements discussed above reduce ASCC's errors
from more than 1 eV to about 0.2 eV on average, which still lags
slightly behind EOM-CCSD's 0.1 eV performance in this category.
For charge transfer states, however,
comparisons to EOM-CCSDT on an expanded test set,
including the tetrafluoroethylene/ethylene dimer and
a number of states from the benchmark of Kozma and coworkers, \cite{kozma2020new}
show that the improved ASCC outperforms EOM-CCSD by about 0.25 eV
to achieve a sub-0.1 eV mean unsigned error.
This accuracy is remarkable for charge transfer states,
which have long vexed leading excited state theories.
To walk readers through how these improved accuracies came about,
we will begin with an overview of ASCC before exploring the
perturbative analysis and the details of the computational results.

\section{Theory}

\subsection{Aufbau Suppression}

In single-reference ground state CC,
\cite{bartlett2007coupled,helgaker2013molecular,crawford2007introduction,shavitt2009many}
the theory is motivated by an exponential ansatz
\begin{equation}
    \ket{\Psi _{\text{CC}}}=e^{\hat{T}}\ket{\phi _0}=\left(1+\hat{T}+\frac{1}{2}\hat{T}^2 + \cdots \right)\ket{\phi _0} \label{eqn: CC equation}
\end{equation}
in which the cluster operator $\hat{T}$ is composed of sums of excitation operators and $\ket{\phi _0}$ is the closed-shell, Aufbau determinant, usually determined via Hartree-Fock (HF) theory. 
\textcolor{black}{
\begin{equation}
    \hat{T}=\hat{T}_1+\hat{T}_2+\cdots+\hat{T}_N
\end{equation}
\begin{equation}
    \hat{T}_N=\left(\frac{1}{N!}\right)^2\sum_{ij\cdots}\sum_{ab\cdots}t_{ij\cdots}^{ab\cdots}\hat{a}^\dagger\hat{b}^\dagger\cdots\hat{j}\hat{i}
\end{equation}
Here, indices $ij\cdots$ represent occupied orbitals while indices $ab\cdots$ represent virtual orbitals, and the operators for these indices represent the usual second quantization creation and annihilation operators.}
If the full cluster operator is utilized, this ansatz can exactly reproduce FCI; however, in practice, it is truncated for computational expediency. One consequence of this truncation is an increased reliance on the choice of the reference determinant. Though CC theory tends to be less sensitive to this choice than CI or perturbation theory,\cite{bartlett2007coupled} it nevertheless requires at least a qualitatively correct reference to produce meaningful results. This is well evidenced by the breakdown of single-reference CC theory in situations involving strong correlation, where a single determinant is no longer a good qualitative representation of the overall wave function.\cite{lyakh2012multireference,kohn2013state}

To retain as much of the framework of single-reference ground
state CC as possible, and to avoid challenges that can arise in
multi-reference CC, ASCC also adopts the Aufbau determinant as
its formal reference.
However, unlike in the ground state theory, this determinant is far
from a good qualitative starting point for an electronically excited state.
Indeed, most excited states will contain $\ket{\phi_0}$ with only
a very small coefficient if at all.
To mold this formal reference into a more appropriate shape, ASCC
therefore incorporates a de-excitation operator $\hat{S}^\dagger$ within a
second exponential.
\begin{equation}
    \ket{\Psi _{\text{ASCC}}}=e^{-\hat{S}^\dagger}e^{\hat{T}}\ket{\phi _0}
\label{eqn: ogascc}
\end{equation}
Within this general form, remarkably simple choices for $\hat{S}$
and $\hat{T}$ lead to qualitatively correct starting points for singly
excited states.
For example, \textcolor{black}{for a state dominated by a single CSF (a 1-CSF state)} in which an electron has been excited
from the ``hole'' orbital $h$ to the ``particle'' orbital $p$, choosing
\begin{equation}
    \hat{S}_{\mathrm{1{\text -}CSF}} = \frac{1}{\sqrt{2}} \left(
        \hat{p}^\dagger_\uparrow \hat{h}_\uparrow + 
        \hat{p}^\dagger_\downarrow \hat{h}_\downarrow
        \right)
\label{eqn:1CSF-S}
\end{equation}
and setting appropriate values \cite{tuckman2024aufbau}
for the singly- and doubly-excited particle/hole
amplitudes within $\hat{T}$ converts Eq.\ (\ref{eqn: ogascc})
into a much more appropriate starting point
\begin{equation}
    \ket{\Psi_{\text{ref}}} = 
    \hat{S}_{\mathrm{1{\text -}CSF}} \ket{\phi_0}
    \label{eqn: 1csf}
\end{equation}
for the excited state in question.
\textcolor{black}{Once this starting point is set, the idea is to hold $\hat{S}$ fixed while optimizing $\hat{T}$ to recover weak correlation.}


To develop these wave function foundations into a fully fledged
CC theory, one follows the same basic steps and arrives at almost
the same equations as one would for ground state CC.
Specifically, one assumes that the wave function is a Hamiltonian
eigenstate,
\begin{equation}
    \hat{H}\ket{\Psi _{\text{ASCC}}}=E\ket{\Psi _{\text{ASCC}}}
    \label{eqn:eigenstate-assumption}
\end{equation}
rearranges the expression
using a similarity transform,
\begin{align}
    e^{-\hat{T}} \bar{H} e^{\hat{T}} \ket{\phi_0} &= E \ket{\phi_0}
    \label{eqn: sim eig} \\
    \bar{H}&=e^{S^\dagger}\hat{H}e^{-S^\dagger}
    \label{eqn: sim}
\end{align}
and projects with the Aufbau $\bra{\phi_0}$ and various excited
determinants $\bra{\mu}$ to produce the energy and amplitude working equations.
\begin{align}
    \bra{\phi_0} e^{-\hat{T}} \bar{H} e^{\hat{T}} \ket{\phi_0} &= E
    \label{eqn:energy} \\
    \bra{\mu} e^{-\hat{T}} \bar{H} e^{\hat{T}} \ket{\phi_0} &= 0
    \label{eqn:amp}
\end{align}
So long as $\hat{S}$ remains a one-body operator, the transformation
in Eq.\ (\ref{eqn: sim}) has at most an $N^5$ cost, and one is
left with the same working equations as one would have if
a ground state approach had employed the same $\hat{T}$.
The key differences are that the one- and two-electron
integral values within $\bar{H}$, which remains two-body,
are different than those in the original $\hat{H}$, and one
may choose the molecular orbital (MO) basis for the Aufbau
determinant from an excited-state-specific theory such as
excited state mean field theory (ESMF).
\cite{shea2020generalized,hardikar2020self}

Thanks to its exponential form and the fact that similarity
transforms preserve the Hamiltonian's spectrum,
ASCC retains the same strong formal properties as ground state CC.
Its ansatz is product separable, its energy is size extensive and
size consistent, and its excitation energies will be size intensive.
Further, it will recover FCI in the limit
of a sufficiently flexible $\hat{T}$, which
makes systematic approaches to selecting the excitations
to include especially desirable.
As in ground state theory, MBPT offers a powerful framework
for \textcolor{black}{determining the relative importance of the different amplitudes}.

\subsection{Truncation of the Cluster Operator}
\label{sec: perb}

Before developing the analysis for ASCC, let us briefly review
how one can use MBPT to motivate specific truncations of $\hat{T}$
in ground state CC.
To start, we separate the Hamiltonian into a zeroth order
\textcolor{black}{piece, which we define as the block diagonal occupied-occupied and virtual-virtual part of the Fock operator},
and a first order piece, which contains everything else,
\textcolor{black}{noting that indices $i$ and $j$ range over all occupied MOs, while indices $a$ and $b$ range over all virtual MOs}.
\begin{align}
    \hat{H}^{(0)}&= \sum_{ij} f_{ij} \hat{i}^\dagger \hat{j} + \sum_{ab} f_{ab} \hat{a}^\dagger \hat{b} \label{eqn: h0} \\
    \hat{H}^{(1)}&= \hat{H} - \hat{H}^{(0)}
\end{align}
\textcolor{black}{We note that while a choice of canonical MOs conveniently produces a purely diagonal Fock operator, this more general form gives $\hat{H}^{(0)}$ orbital invariance with respect to occupied-occupied and virtual-virtual rotations,\cite{bartlett2007coupled,shavitt2009many} which can be particularly helpful when localizing some or all of the orbitals for local correlation purposes.} 
With this division of the Hamiltonian, one can straightforwardly
motivate CCSD as a good lowest-order choice for the ground state theory.
To do so, one notes that the only terms that make first order
contributions to any of the amplitude equations are
\begin{align}
    Z_i^a &\equiv \bra{\phi _{i}^{a}}\hat{H}^{(1)}\ket{\phi _0}
    \label{eqn: perb singles}
\end{align}
and
\begin{align}
    Z_{ij}^{ab} &\equiv \bra{\phi _{ij}^{ab}}\hat{H}^{(1)}\ket{\phi _0}
    \label{eqn: perb doubles}
\end{align}
which contribute to the singles and doubles equations, respectively.
Thus, the singles and doubles are identified as the lowest order
parts of $\hat{T}$, and the MBPT motivation for CCSD is achieved.
Of course, if HF is used as the reference, then $Z_i^a$ would be zero
due to Brillouin's theorem, but the singles are typically included
anyways as doing so can be accomplished without adding any
new $O(N^6)$ terms to the theory, and they allow the
orbitals to relax in the presence of correlation.
\cite{thouless1960stability}

Unlike the ground state theory, in which all parts of $\hat{T}$ are
first order or smaller, Aufbau suppression necessitates that
$\hat{S}$ and at least a handful of amplitudes within $\hat{T}$
be zeroth order.
\textcolor{black}{To help delineate which amplitudes are zeroth order and to group amplitudes according to their commutative properties, we rearrange the ASCC wave function as follows.}
\begin{equation}
    \ket{\Psi _{\text{ASCC}}}=e^{\hat{T}^N}e^{-\hat{S}^\dagger}e^{\hat{T}^M}e^{\hat{T}^P}\ket{\phi _0}
    \label{eqn: pnm split}
\end{equation}
Here, $\hat{T}^P$ contains what we will refer to as the
primary amplitudes, which
have only \textcolor{black}{the primary} hole and particle indices
and, \textcolor{black}{through their interaction with $\hat{S}^\dagger$,}
transform the formal reference into the qualitatively
correct excited state reference.\cite{tuckman2024aufbau}
\textcolor{black}{On the other hand, $\hat{T}^N$ contains amplitudes with no primary indices, and, as a result, commutes with $\hat{S}^\dagger$. Finally, $\hat{T}^M$ contains ``mixed'' amplitudes which contain both primary and nonprimary indices, and therefore do not commute with $\hat{S}^\dagger$. Though this partitioning of the amplitudes differs slightly from our previous work,\cite{tuckman2024aufbau} the strength of this additional subdivision will be made clear in the upcoming perturbative analysis.}

\textcolor{black}{With the cluster amplitudes partitioned, it is now straightforward to demonstrate that all of the zeroth order cluster operators reside in $\hat{T}^P$.}
For example, in the single-CSF case,
Eq.\ (\ref{eqn: ogascc}) is converted into Eq.\ (\ref{eqn: 1csf})
by setting $\hat{T}^M = \hat{T}^N = 0$ and initializing $\hat{T}^P$ to
\begin{equation}
    \label{eqn:TP-guess}
    \hat{T}^{P(0)}
    =
    \beta~\hat{S}
    + \gamma~\hat{S}^2
\end{equation}
in which $\beta = 1$, $\gamma = -1/2$, and
$\hat{S}$ is taken from Eq.\ (\ref{eqn:1CSF-S}).
In all states, we define the primary orbitals as those whose indices appear within $\hat{S}$.
There are two primary orbitals ($h$ and $p$) in a 1-CSF state, four in a 2-CSF state,
six in a 3-CSF state, and so forth.

\textcolor{black}{The presence of zeroth order amplitudes must be handled carefully when partitioning the Hamiltonian for perturbation
theory.
Were we to follow the ground state partitioning of Eq.\ (\ref{eqn: h0}), the off-diagonal part of $\hat{H}^{(0)}$ could interact with $\hat{T}^P$ to create zeroth order contributions to the amplitude equations for $\hat{T}^N$ and $\hat{T}^M$, at which point these pieces, which were not necessary for constructing the reference, would perversely be labeled as zeroth order.  To avoid this issue in ASCC, we instead choose a partition that leaves $\hat{H}^{(0)}$ with four block diagonal pieces: the primary hole orbitals block, the nonprimary occupied orbitals block, the primary particle orbitals block, and the nonprimary virtual orbitals block. We again use indices $h$ and $p$ for the primary hole and particle orbitals respectively, but note that now $i$ and $j$ refer to only nonprimary occupied orbitals while $a$ and $b$ refer to only nonprimary virtual orbitals.}

\begin{align}
    \hat{H}^{(0)}&= \sum_{h_1h_2} f_{h_1h_2} \hat{h}_1^\dagger \hat{h}_2 + \sum_{ij} f_{ij} \hat{i}^\dagger \hat{j} \notag\\ &\quad+\sum_{p_1p_2} f_{p_1p_2} \hat{p}_1^\dagger \hat{p}_2+ \sum_{ab} f_{ab} \hat{a}^\dagger \hat{b} \label{eqn: ascch0} \\
    \hat{H}^{(1)}&= \hat{H} - \hat{H}^{(0)}
\end{align}

\textcolor{black}{By partitioning the Hamiltonian in this way, no zeroth order contributions to $\hat{T}^N$ or $\hat{T}^M$ arise from the interaction of the zeroth order Hamiltonian with $\hat{T}^{P(0)}$.
This decoupling also means that ASCC is only invariant to orbital rotations within each of the four orbital blocks, but this result is somewhat expected as this invariance pattern matches the invariance of the singly-excited reference itself.  Put another way, ASCC has the same orbital invariance relationship with its reference as we see in the ground state; the CC theory is invariant to the same orbital rotations as the reference is invariant to.
Notably, semicanonicalization
\cite{handy1989size,bartlett2007coupled,shavitt2009many} of the nonprimary orbitals remains possible, and so expressions can be simplified by making the blocks in Eq.\ (\ref{eqn: ascch0}) diagonal.  ASCC is also invariant to localization transformations within the non-primary blocks, which may in future support local correlation approaches.}

\textcolor{black}{With this Hamiltonian partitioning chosen, we now find that the largest components of $\hat{T}^N$ and $\hat{T}^M$ are first order.} Because $\hat{T}^N$ commutes with $\hat{S}^\dagger$, it acts much like a multi-reference CC cluster operator in that
it operates on the already-constructed multi-determinant
excited state reference.
As such, its perturbative analysis will follow the familiar
ground state pattern, with singles and doubles being first order,
triples being second order, and so on.
In contrast, $\hat{T}^M$ does not commute with $\hat{S}^\dagger$,
and consequently some of the amplitudes within $\hat{T}^M$
will be lower order than in the ground state theory.

To determine which amplitudes in $\hat{T}^M$ will be
first order, we begin by noting that
\begin{align}
   \bar{H}^{(0)} & = e^{S^\dagger} \hat{H}^{(0)} e^{-S^\dagger}
\end{align}
is a one-body operator whose \textcolor{black}{occupied-virtual} off-diagonal component
has only primary indices and is purely de-exciting
(this follows because $\hat{H}^{(0)}$ is \textcolor{black}{block} diagonal
and because $\hat{S}^\dagger$ is purely de-exciting
and has only primary indices).
As a consequence, although $\bar{H}^{(0)}$ is not
\textcolor{black}{block} diagonal, terms like
\begin{equation}
    \bra{\mu} \left[
    \bar{H}^{(0)}, \hat{T}^{(1)}
    \right] \ket{\phi_0}
\end{equation}
are only nonzero for determinants $\ket{\mu}$
that are equally or less excited than the amplitudes
already present in $\hat{T}^{(1)}$.
We therefore do not need to worry about them for
identifying first order amplitudes, although
they will become important later when considering
how strongly specific amplitudes influence the energy.
For finding which amplitudes will be first order,
it is instead commutators between the first order
\begin{align}
   \bar{H}^{(1)} & = e^{S^\dagger} \hat{H}^{(1)} e^{-S^\dagger}
\end{align}
and the zeroth order $\hat{T}^P$ that cause a subset of
more highly excited mixed amplitudes to join the singles and
doubles within $\hat{T}^{(1)}$.
Narrowing our focus to these commutators, let us first
complete the analysis for the single-CSF case before
showing how to generalize it to multi-CSF states.

\subsubsection{Single-CSF ASCC}

In a single-CSF state, there is only one doubly excited
amplitude within the zeroth order cluster operator in Eq.\ (\ref{eqn:TP-guess}).
Thus, the only terms that
lead to first order amplitudes with more than
two excitations are the terms within
\begin{equation}
    \label{eqn:H1TP}
    \bra{\mu}\left[
    \bar{H}^{(1)}, \hat{T}^{P(0)}
    \right]\ket{\phi_0}
\end{equation}
in which this all-primary double participates in a
single contraction with the two-electron part of $\bar{H}^{(1)}$.
The resulting triple excitation will have at least three
primary indices --- as only one of the all-primary double's
indices was contracted away --- leading to two possibilities.
\begin{align}
    \bra{\phi _{h \bar{h} k}^{p b c}}\left[\bar{H}^{(1)},\hat{T}^{P(0)}\right]\ket{\phi _0} &\neq 0\\
    \bra{\phi _{h j k}^{p \bar{p} c}}\left[\bar{H}^{(1)},\hat{T}^{P(0)}\right]\ket{\phi _0} &\neq 0
\end{align}
Note we use the $\bar{p}$ and $\bar{h}$ notation to denote
the opposite-spin particle or hole orbital, respectively.
We therefore see that, in addition to the singles and doubles,
ASCC's first order cluster operator contains this relatively
small $O(o^2v+ov^2)$ slice of mixed triples, which we will
write as $\hat{T}^M_{3'}$.
Interestingly, this gives the same ansatz as we used previously
for single-CSF ASCC, which we had originally motivated by
arguing that one should include all single and double excitations
relative to the excited state reference. \cite{tuckman2024aufbau}
Thus, as in our previous work, single-CSF excitation energy
calculations employ the following cluster operators for the ground and excited states, respectively.
\begin{align}
    \hat{T}_{\text{CCSD}}^{\text{1-CSF}}&=\hat{T}_1+\hat{T}_2\\
    \hat{T}_{\text{ASCCSD}}^{\text{1-CSF}}&=\hat{T}_1+\hat{T}_2+\hat{T}^M_{3'}
\end{align}
While the inclusion of these triples may at first glance raise
computational efficiency concerns, the fact that they
have at least three primary indices
\textcolor{black}{(which range over only $O(1)$ values)}
causes the cost of the worst new terms to scale as only $O(N^5)$.
\textcolor{black}{This is most readily seen by noting that a full CCSDT implementation has at worse $O(N^8)$ scaling terms involving the triples, which upon restricting three of the indices' ranges to be $O(1)$ have their scaling reduced to $O(N^5)$ or less.}
As a result, the $O(N^6)$ parts of ASCC remain exactly the same
as in CCSD, as intended.

With the first order wave function determined, one may naturally
wonder to what order the ASCC energy will be correct to.
Though a full perturbative analysis of the ASCC energy is beyond
the scope of this study, we will point out that it is only complete
through second order as a result of the \textcolor{black}{occupied-virtual} off-diagonal
elements in $\bar{H}^{(0)}$
(see Section \ref{sec: linearization} below for examples
of missing third order contributions).
This property contrasts with CCSD, whose energy is complete through
third order.
To get the ASCC energy complete through third order, we would
need to include amplitudes like the
$O(o^2v^2)$ slice of mixed triples (and mixed quadruples)
that contain exactly one (two) particle-hole primary index pair.
Though these slices are relatively small compared to the full
triples and quadruples, their inclusion would result in
additional $O(N^6)$ terms which, in the present study at least,
we are looking to avoid.
It is worth noting, however, that because these amplitudes must
contain hole and particle indices, they would be expected to
be small in cases where one or more of the remaining indices
referred to an orbital spatially distant from the hole
and particle orbitals.
Therefore, we would expect the energetic correction from these
terms to be local in nature, and in the future it may be
possible to make this correction at lower cost using
local correlation techniques.
In the present study, however, we simply exclude these
amplitudes in order to maintain strict parity with CCSD's
$O(N^6)$ terms, and so our energies are only complete
through second order.

\subsubsection{Two-CSF ASCC}
\label{sec: multicsf}

Suppose now that the excited state in question contains two CSFs with large coefficients, leading us to aim for an excited state reference of the form
\begin{equation}
    \ket{\Psi_{\text{ref}}}=a\left(\ket{\phi _{h_1} ^{p_1}}+\ket{\phi _{\bar{h}_1}^{\bar{p}_1}}\right)+b\left(\ket{\phi _{h_2} ^{p_2}}+\ket{\phi _{\bar{h}_2}^{\bar{p}_2}}\right), \label{eqn: 2csf}
\end{equation}
in which normalization implies that $2a^2+2b^2=1$.
Though the goal of Aufbau suppression can now be achieved in a number of different ways, here we follow the same path
as in the single-CSF case and define
$\hat{S}=\hat{T}^{P(0)}_1$ as the one-body operator that,
when acted on $\ket{\phi_0}$, produces $\ket{\Psi_{\text{ref}}}$.
The key difference from the 1-CSF case is that \textcolor{black}{$\hat{T}^{P(0)}$}
now requires the all-primary triples and quadruples, in
addition to the all-primary doubles, in order to
eliminate higher order terms and ensure that the
initial guess put into Eq.\ (\ref{eqn: pnm split})
yields Eq.\ (\ref{eqn: 2csf}) as the starting point.

Previously,\cite{tuckman2024aufbau} we opted for a minimalist
fleshing out of the rest of the 2-CSF $\hat{T}$ operator,
and in particular what we now denote as $\hat{T}^M$,
by including only the singles, the doubles, and the
the $\hat{T}^M_{3'}$-style slices of the
triples related to each of the two primary CSFs.
The accuracy of this approach was quite unsatisfactory, with
excitation energy errors of about 1 eV strongly hinting
that important pieces of the wave function were missing.
Turning now to a perturbative analysis of what parts of
$\hat{T}$ will be first order in the 2-CSF case, we see that
key amplitudes had indeed been overlooked.

As in the 1-CSF case, terms of the form seen in Eq.\ (\ref{eqn:H1TP})
lead to first order amplitudes beyond just the singles and doubles.
However, as the zeroth order $\hat{T}^P$ now contains all-primary
triple and quadruple excitations, we get cubic-sized slices of
the triples, quadruples, and quintuples
appearing at first order.
Specifically, the terms
\begin{align}
    \bra{\phi _{h_1 \bar{h}_1 h_2 l}^{p_1 \bar{p_1} c\phantom{_2} d}}\left[\bar{H}^{(1)},\hat{T}_3^{P(0)}\right]\ket{\phi _0} &\neq 0\\
    \bra{\phi _{h_1 \bar{h}_1 k\phantom{_2} l}^{p_1 \bar{p}_1 p_2 d}}\left[\bar{H}^{(1)},\hat{T}_3^{P(0)}\right]\ket{\phi _0} &\neq 0\\
    \bra{\phi _{h_1 \bar{h}_1 h_2 \bar{h}_2 m}^{p_1 \bar{p_1} p_2 d\phantom{_2}  e}}\left[\bar{H}^{(1)},\hat{T}_4^{P(0)}\right]\ket{\phi _0} &\neq 0\\
    \bra{\phi _{h_1 \bar{h}_1 h_2 l\phantom{_2} m}^{p_1 \bar{p}_1 p_2 \bar{p}_2 e}}\left[\bar{H}^{(1)},\hat{T}_4^{P(0)}\right]\ket{\phi _0} &\neq 0
\end{align}
imply that $\hat{T}^{(1)}$ should include the $O(o^2v+ov^2)$ slices 
of quadruples and quintuples that have five and seven primary indices,
respectively.
Crucially, the new terms that arise from including these amplitudes
all scale as $O(N^5)$ or lower, and so they do not interfere with
our goal of adding no new $O(N^6)$ terms.

These extra amplitudes do, however, raise concerns about balance.
While some of them are \textcolor{black}{undoubtedly} describing correlation details
directly related to the excitation, it is hard to imagine that these
broad swaths of amplitudes are not also improving the correlation
treatment of electrons near the primary orbitals in ways that have
nothing to do with the excitation.
One way to get a handle on how biasing this effect might be is to
ask how much correlation energy the analogous ground state amplitudes
(triples and quadruples with three and five primary indices, respectively)
would impart if included atop CCSD.
The (T) correction in molecules like N$_2$, formaldehyde, and acetone
--- whose size might be taken as a crude simulacrum for the collection of
electrons ``near'' the primary orbitals in a larger molecule ---
is about 0.3 to 0.6 eV, and so even if these terms only captured
part of that effect it could seriously impede our accuracy.
Indeed, explicit tests show that enabling these amplitudes often shifts
the ground state energy by about 0.2 eV, and so although they are
clearly not part of $\hat{T}^{(1)}$, we include them in the ground
state calculation when predicting an excitation energy to help
balance the presence of the new pieces of $\hat{T}^{(1)}$ in
the excited state, leading us to the following 2-CSF amplitude choices.
\begin{align}
    \hat{T}_{\text{CCSD}}^{\text{2-CSF}}&=\hat{T}_1+\hat{T}_2+\hat{T}^M_{3'}+\hat{T}^M_{4'}\\
    \hat{T}_{\text{ASCCSD}}^{\text{2-CSF}}&=\hat{T}_1+\hat{T}_2+\hat{T}^M_{3'}+\hat{T}^M_{4'}+\hat{T}^M_{5'}
\end{align}
Note that, in order for the primed slices to have the same meaning
in both states, we perform the ground state calculation in the MO
basis in which occupied-occupied and virtual-virtual rotations
have been applied to bring the two hole and two particle orbitals
into maximum overlap with their excited state ESMF counterparts,
with the remaining occupied and virtual orbitals canonicalized.
For 3-CSF states, the story would be similar, extending to
the cubic slices $\hat{T}^M_{6'}$ and $\hat{T}^M_{7'}$ that
had nine and eleven primary indices.
This would still not add anything worse than more $O(N^5)$ terms
to the theory, but the number of such terms would be substantial, and so
in the present study we have limited our testing to 1- and 2-CSF states.

Just as in the 1-CSF case, our 2-CSF ASCC energy is only complete
through second order.
To make it complete through third order,
$O(o^2v^2)$ slices of the triples, quadruples, quintuples, and hextuples
appear to be necessary.
As in 1-CSF ASCC, adding these would introduce additional
$O(N^6)$ terms, and so we opt not to in this study.
However, as we now appear to be leaving out a larger chunk
of the third order energy contributions than in the 1-CSF case,
we might expect the consequences to be more substantial.
Indeed, the results below show that although the present approach
to 2-CSF states is much more accurate than our previous approach,
its accuracy still lags slightly behind what we get for 1-CSF states.
As discussed above, we are optimistic that future work will be
able to close this gap without meaningfully increasing cost
by exploiting the fact that the amplitudes in question bear
large numbers of primary indices and so should add correlation
that is local in nature.

\subsection{Partial Linearization}
\label{sec: linearization}

Although the \textcolor{black}{occupied-virtual} off-diagonal nature of $\bar{H}^{(0)}$ does not affect which
amplitudes appear within $\hat{T}^{(1)}$, it does have important
consequences for third order energy contributions.
Through terms of the form
\begin{equation}
    \label{eqn:down-ladder}
    \bra{\mu}
    \left[ \bar{H}^{(0)}, \hat{T} \right]
    \ket{\phi_0},
\end{equation}
the off-diagonal of $\bar{H}^{(0)}$,
which de-excites in the primary space,
creates a ``downward ladder'' effect allowing amplitudes
within $\hat{T}^P$ and $\hat{T}^M$ to contribute to the
amplitude equations of less excited amplitudes at their
own order in perturbation theory.
In 1-CSF ASCC, for example, mixed quadruples
show up in the second order part of $\hat{T}^M$ and
make via Eq.\ (\ref{eqn:down-ladder}) a second order
contribution to the mixed triples, which in turn makes a second
order contribution to the doubles via another downward ladder term.
This second order contribution to the doubles means that the
energy receives a third order contribution ultimately coming
from $\hat{T}_4^M$.
In contrast, ground state CC's largest energy contribution
from $\hat{T}_4$ is fifth order.

As best we can tell, this surprisingly low order contribution from quadruples
shows up in ASCC in order to counteract other side effects of
downward ladder terms.
For example, in 1-CSF ASCC, there is only one choice for
$\beta$ and $\gamma$ in Eq.\ (\ref{eqn:TP-guess}) that exactly
zeros out the Aufbau determinant and the all-primary double excitation.
We use this choice in our initial guess for the amplitudes and,
for an excited state of a different irreducible representation
than the ground state, any move away from this choice would produce
a symmetry violation by reintroducing the Aufbau and/or the double.
Keeping that in mind, note that the term
\begin{equation}
    \label{eqn:T2P-violation}
    \bra{\mu} \left[
    \left[ \bar{H}^{(1)}, \hat{T}_2^M \right], \hat{T}_2^M
    \right] \ket{\phi_0}
\end{equation}
makes a third order contribution to $\hat{T}_2^P$, which is to
say that it changes the value of $\gamma$ and thus violates symmetry.
Due to the downward ladder effect, this change to $\hat{T}_2^P$
leads in turn to a third order change in $\hat{T}_1^P$,
which furthers the symmetry violation and, via
\begin{equation}
    \label{eqn:T1P-3rd-order-trouble}
    \bra{\phi_0}
    \left[ \bar{H}^{(0)}, \hat{T}_1^P \right]
    \ket{\phi_0},
\end{equation}
makes a corresponding third order contribution to the energy.
Since any movement in $\hat{T}_1^P$ or $\hat{T}_2^P$ is a
symmetry violation when the excited state is of different
symmetry than the ground state, the contributions to these
operators at each order of perturbation theory
above zeroth order should cancel each other out, order by order.
Without the $O(o^2v^2)$ slice of the mixed quadruples that lives
in the second order part of $\hat{T}$ and makes a third
order contribution to $\hat{T}_2^P$ via
\begin{equation}
    \bra{\mu} 
    \left[ \bar{H}^{(1)}, \hat{T}_4^M \right]
    \ket{\phi_0},
\end{equation}
the cancellation at third order is disrupted,
resulting in a symmetry violation with
third order energy consequences.

Although it is not as clearly an error as in the case of
symmetry violations, some contributions to $\hat{T}_2^N$
involving the downward ladder also seem to be
unwelcome side effects of Aufbau suppression.
Specifically, the term
\begin{equation}
    \label{eqn:T2-squared-trouble}
    \bra{\mu} \left[
    \left[ \bar{H}^{(0)}, \hat{T}_2^M \right], \hat{T}_2^M
    \right] \ket{\phi_0}
\end{equation}
makes a second order contribution to $\hat{T}_3^M$, which via
the downward ladder leads to a second order contribution to
$\hat{T}_2^N$ and a corresponding third order energy contribution.
Again, we see a $(\hat{T}_2^M)^2$ term, which would correspond
to quadruply excited determinants in the wave function expansion,
making a surprisingly low-order contribution to the energy, this
time via doubles amplitudes whose indices are all non-primary
and so have no direct connection to the excitation.
Compared to the symmetry argument above, it is harder to be sure
that this case is a purely erroneous side effect of Aufbau
suppression that should be counteracted by the mixed
quadruples (by repeated downward laddering
from $\hat{T}_4^M$ to $\hat{T}_2^N$),
but seeing $(\hat{T}_2)^2$ terms make third order energy
contributions makes us quite suspicious that this is basically
what is going on.
Thus, we can identify multiple issues that help explain why
the energy contains a third order contribution from
$\hat{T}_4^M$, but they all seem to be cases of
$\hat{T}_4^M$ cleaning up a mess made by terms containing
nonlinear powers of mixed amplitudes.

If this cleanup job is indeed the leading-order role of $\hat{T}_4^M$,
then we may be able to inexpensively capture the accuracy improvements
that it would bring by instead dropping nonlinear
terms like like Eq.\ (\ref{eqn:T2P-violation})
and Eq.\ (\ref{eqn:T2-squared-trouble})
in a partial linearization of the theory.
Which terms should go?
As the off-diagonal of $\bar{H}^{(0)}$ only de-excites
in the primaries, the downward ladder effect is not
an issue for fully non-primary amplitudes.
For fully primary amplitudes, the nonlinear terms are
crucial for converting the formal reference into the
excited state reference.
Thus, our partial linearization will only discard
terms containing two or more powers of $\hat{T}^M$.
Further, because nonlinear powers of $\hat{T}_1$ are important
for relaxing the orbitals in the presence of correlation,
\cite{thouless1960stability}
we will ignore $\hat{T}^M_1$ when counting up powers
of $\hat{T}^M$.
To help ensure balance, we apply the same term removal
to the ground state equations.
The resulting partially linearized ASCC (PLASCC) theory
thus introduces no new amplitudes and continues to have
exactly the same $O(N^6)$ terms as CCSD.
Further, as it differs from ASCC only by the removal
of connected diagrams, PLASCC remains size consistent
and extensive for absolute energies and intensive
for excitation energies.
\cite{goldstone1957derivation,bartlett1978many,shavitt2009many}
Although its energy will remain complete only through second order,
we hope that, by sidestepping the worst consequences of ASCC's
downward $\bar{H}^{(0)}$ ladder, it will be more accurate.

\subsection{A second ansatz}

Although we did not realize this in our original study
of ASCC, we have since found that for excited states in which the Aufbau
determinant makes a small but nonzero contribution,
there are two similar but distinct ansatz choices for ASCC.
To see why, start by writing the desired excited state
reference wave function using a normalization convention
in which the coefficient on the
singly excited $\hat{S}\ket{\phi_0}$ part is one.
\begin{equation}
    \ket{ \Psi_{ \text{ref} } } =
    \alpha \ket{\phi_0} + \hat{S}\ket{\phi_0}
    \label{eqn:ref-with-nz-aufbau}
\end{equation}
To hit this target using Eq.\ (\ref{eqn: pnm split})
when $\alpha$ is small but not zero, we can
update the $\beta$ and $\gamma$ values in  Eq.\ (\ref{eqn:TP-guess})
so that the Aufbau suppression is no longer perfect.
Choosing $\gamma=-\beta^2/2$ to eliminate the doubly excited
part of the expansion, setting $\hat{T}^N=\hat{T}^M=0$,
and applying our normalization condition, we get
\begin{equation}
    \ket{\Psi_{\text{ASCC}}} \rightarrow
    \frac{1 - \beta}{\beta} \ket{\phi_0}
    + \hat{S}\ket{\phi_0}.
\end{equation}
As before, if we choose $\beta=1$, the Aufbau component
is completely eliminated, but that is no longer what we want.
To get an $\alpha$ value slightly above zero, we would choose
$\beta$ slightly below one, and for an $\alpha$ value slightly
below zero, we would choose $\beta$ slightly above one.

So far, this looks like just one ansatz, but consider what
happens if we flip the sign of the hole orbital.
The Aufbau determinant, containing two copies of this orbital,
keeps its original sign, but the singly excited determinants,
with only one copy of this orbital, change sign.
Dividing through by this sign to maintain the normalization
convention, we see that by playing with the sign of the hole orbital
(or by similar logic the sign of the particle orbital)
we are free to flip the sign of $\alpha$.
Thus, we can hit our desired target for the excited state
reference using a value of $\beta$ that is either slightly
above one or slightly below one, as long as the
orbital signs are set accordingly.
For the purpose of constructing the excited state reference
in Eq.\ (\ref{eqn:ref-with-nz-aufbau}),
this choice does not make a difference; either way we
get the job done.
However, it does  make a difference in the amplitude
equations, especially for terms with nonlinear powers of $\hat{T}^P$
in which the differences between
$\beta>1$ and $\beta<1$ will be emphasized.
Thus, in excited states where the Aufbau determinant is expected
to make a small but nonzero contribution, there turn out to be
two subtly different choices for the ASCC ansatz.

In the present study, we avoid the choice entirely by
evaluating the energy for both the ``small $\beta$''
and the ``large $\beta$'' cases and averaging them to produce
the reported ASCC energy.
In states where point group symmetry forces the Aufbau
coefficient to be zero, this approach makes no difference,
as the small and large cases both simplify
into the $\beta=1$ scenario.
In other states, we really do get two different energies,
but they are typically separated by
just a tenth of an eV or so,
and, as we will see in the results, the accuracy of their
average proves to be quite good.
In the future, it will be interesting to investigate more closely
whether there is a strong reason to favor one case over the other,
but for now we avoid the question via averaging.
\section{Results}
\label{sec:results}

\subsection{Computational Details} \label{sec: comp}
EOM-CCSD calculations were performed using PySCF,\cite{sun2015libcint,sun2018pyscf,sun2020recent} while reference calculations with CCSD(T), EOM-CCSDT, and \textcolor{black}{LR}-CC3 were performed with Q-Chem\cite{epifanovsky2021software} and PSI4\cite{smith2020psi4} respectively. Complete active space self-consistent field (CASSCF) calculations were performed using Molpro.\cite{MOLPRO} In contrast to both the calculated and literature reference values, the ASCC and EOM-CCSD calculations were performed without the frozen core approximation. Though this is expected to make a small difference to the excitation energies ($\sim$0.02eV),\cite{loos2018mountaineering,loos2020mountaineering} the high accuracy frozen-core results nevertheless serve as excellent reference values. All single-CSF and two-CSF tests were performed in the aug-cc-pVDZ basis while the charge transfer tests were performed in the cc-pVDZ basis in order to match their respective reference methods in the literature. All geometries come from their respective studies, with the exception of the ammonia-difluorine system, which was further separated to an intermolecular distance of 5 \r{A} due to ESMF experiencing significant root mixing at shorter distances. This new geometry, along with the newly added 3,5-difluoro-penta-2,4,dienamine molecule, may be found in the Supporting Information. For these new geometries, we calculate the reference energy with EOM-CCSDT for the ammonia-difluorine system and \textcolor{black}{LR}-CC3 for 3,5-difluoro-penta-2,4,dienamine due to its larger size.

For ASCC, the excited state reference was determined via the excited state mean field (ESMF) method,\cite{shea2018communication,shea2020generalized,hardikar2020self} which can essentially be summarized as a CIS wave function with orbital relaxations. 
Any CSF with a singular value greater than 0.2 in the ESMF wave function was considered part of the ASCC reference. Convergence in ASCC was considered achieved when both the maximum residual was no greater than $10^{-5}$ and the energy changed by less than $10^{-7}$ Ha in an iteration. On rare occasions, ASCC or PLASCC would experience convergence issues, particularly when the ESMF wave function had significant deficiencies as compared to EOM-CCSD. If neither of the two solutions would converge, these states were entirely removed for all methods reported. However, on a few states only one of two ASCC or PLASCC solutions from the two separate ansatz would fail to converge. For these particular states, which are marked in the Supporting Information, the energy reported is not the arithmetic mean of the two solutions, but rather just the energy of the only converged solution. Furthermore, for a few states ASCC or PLASCC would stall near, but not quite at, the convergence criterion. Because the energy was changing below the level of energetic precision reported, these states were still included, though they are marked in the Supporting Information.  Though an improved approximate Jacobian may help to alleviate these issues in the future, it is also possible that an investigation of more robust reference methods may also help to avoid these issues entirely. Nevertheless, removal of these states does not have a significant effect on the statistics reported. Default tolerances were used for all other methods.

\subsection{Implementation and Timing Analysis}

The current iteration of the ASCC code is entirely factorized and implemented via automated code generation through a scheme which takes inspiration from that of Kallay and coworkers.\cite{kallay2001higher} A brief explanation of the overarching ideas that the autogenerator implements may be found in the Supporting Information, though we also highlight some of the most important conclusions here. The automated factorization which takes place ensures that tensor contractions occur in their optimal order for each individual term, but as a result of the emphasis on a term by term evaluation, misses more global intermediates such as the well known $\tau _{ij}^{ab}=t_{ij}^{ab}+t_i^at_j^b$ intermediate found in many hand factorized codes. Nevertheless, this factorization scheme is guaranteed to produce the correct asymptotic scaling, even if it does not produce the factorization with the theoretical fewest FLOPs. The current implementation also stores the cluster amplitudes in their unrestricted, antisymmetrized forms. While this has the advantage of a more vectorizable implementation, which is favorable for achieving processor efficiency in matrix multiplication as well as simplicity in the implementation, it has the disadvantage of introducing significant redundancy and sparsity in the cluster amplitude tensors, especially for the higher excitations present in the 2-CSF implementation. For example, in the 2-CSF $T_4 ^P$ operator there is exactly one unique, nonzero amplitude. However, in the current implementation an extremely sparse, 256 element tensor is stored and calculated in the residual equations. Though this effect is also present in the 1-CSF implementation, because triples are the highest excited cluster operators included the computational storage is at worst affected by a factor of 2, which is a sharp juxtaposition to the larger impact in 2-CSF.  Finally, for computational expediency this implementation also makes use of spin adaptation whose details may also be found in the Supporting Information. 

\begin{figure}
    \centering
    \includegraphics[width=\linewidth]{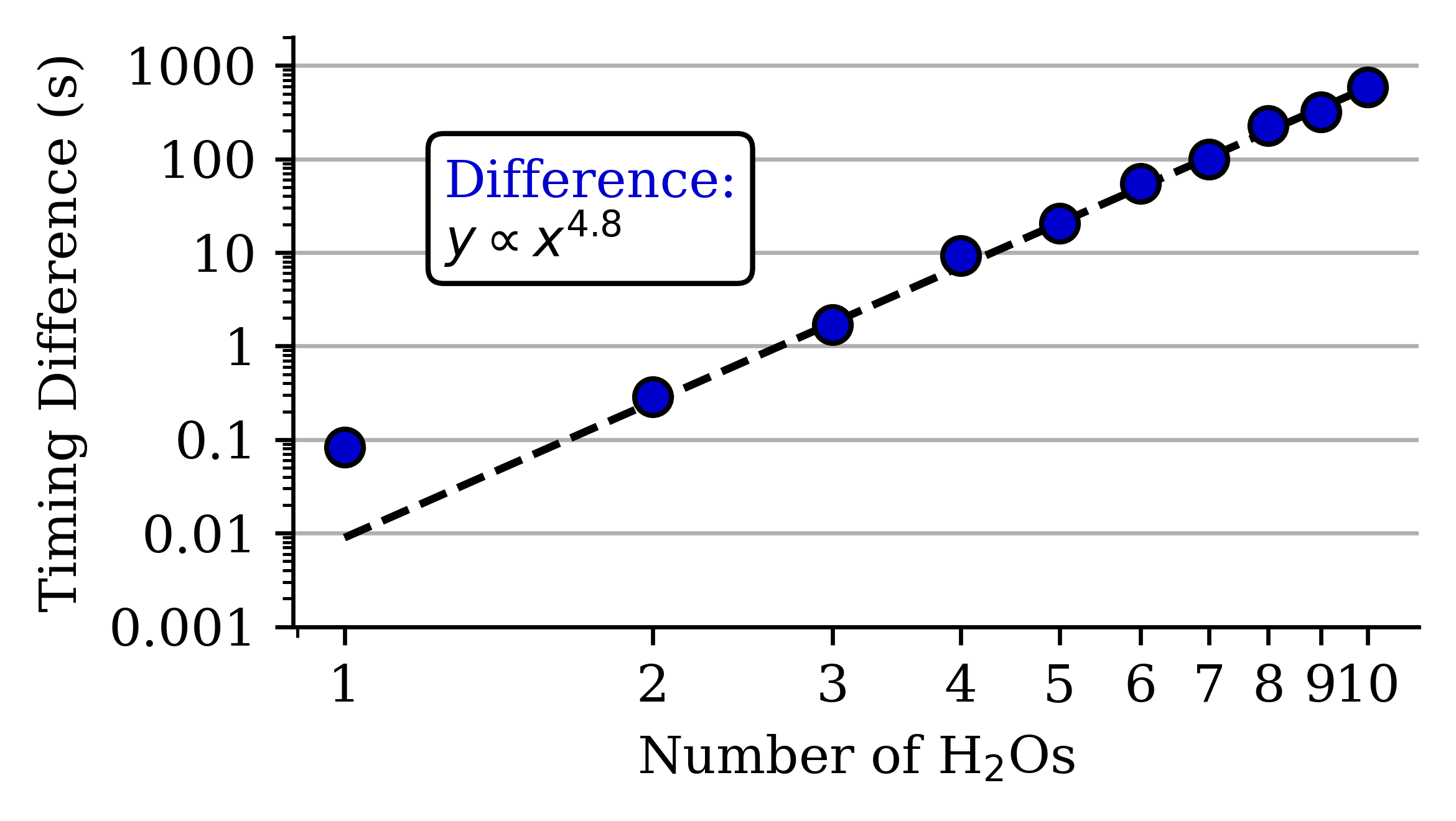}
    \caption{Wall time differences per iteration between single-CSF ASCCSD and \textcolor{black}{ground state }CCSD codes generated via the same autogenerator
    for varying numbers of water molecules in a cc-pVDZ basis.
    Both methods were run on a single core of an Intel
    Xeon Gold 6330 2.0 GHz processor.
    Dashed line shows $y = a x^m$ function of best fit
    for the final five points. }
    \label{fig: timing}
\end{figure}

As per our design goal, ASCCSD's highest scaling terms are $O(N^6)$,
and these terms are the same $O(N^6)$ terms that appear in
traditional CCSD.
Figure \ref{fig: timing} shows the per-iteration timing difference
between the single-CSF ASCCSD and \textcolor{black}{ground state }CCSD methods (when implemented
using the same autogenerator) as a function of the number of
water molecules in a cc-pVDZ basis.
Though not shown, results for PLASCCSD are almost identical
to those of ASCCSD.
As can be seen, the difference between the ASCCSD and CCSD methods scales as no more than $O(N^5)$, corroborating the fact that the $O(N^6)$ terms are identical in the two theories. Therefore, while more meticulous software engineering which takes into account the factorization and spare tensor considerations mentioned above may improve the timing difference in small systems, this implementation already achieves an important computational parity between ASCCSD and CCSD in the large-system limit.

\textcolor{black}{Finally, the last point to consider when comparing the timing of ASCCSD with traditional CCSD is the speed of convergence. At present, little attention has been given to optimizing the iterative scheme of ASCCSD, as it still utilizes the same diagonal approximate Jacobian update scheme augmented with DIIS typically found in many ground state CCSD codes, as outlined in our previous work.\cite{tuckman2024aufbau} Nevertheless, ASCCSD typically converges to the tight convergence criteria outlined in Section \ref{sec: comp} within approximately 20-40 iterations with an average of 27 iterations. Although this is slower than the approximately 10-20 with an average of 13 iterations required by our implementation of CCSD on these states, the perturbative framework outlined above offers a promising path towards improving the iterative convergence of ASCC in future by identifying and including key off-diagonal terms in the approximate Jacobian.}

\subsection{Benchmark on Valence \textcolor{black}{and Rydberg} Excitations}

\begin{figure*}
    \centering
    \includegraphics[width=\linewidth]{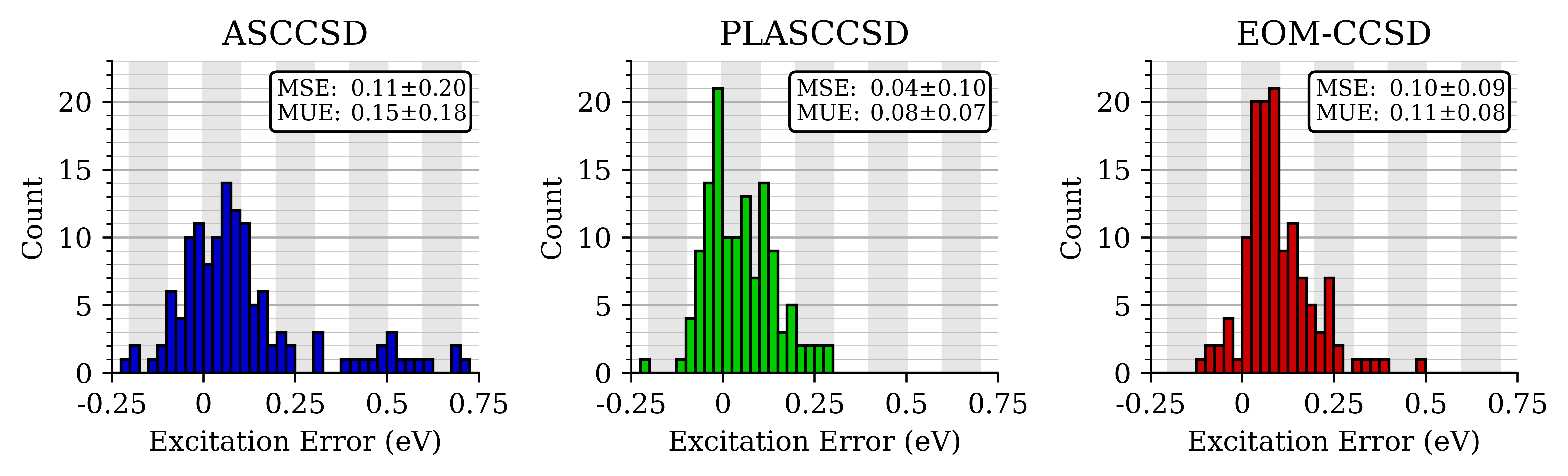}
    \caption{Histograms of the excitation energy error distributions for ASCCSD (left), PLASCCSD (center), and EOM-CCSD (right) for 130 single-CSF states in an aug-cc-pVDZ basis, with reference values of at least CCSDT quality. ASCCSD and PLASCCSD each have one value not shown, with errors of 0.90 and -0.42 respectively.
    The mean signed error (MSE) and mean unsigned error (MUE) are shown at
    the top of each plot \textcolor{black}{along with their standard deviations}.}
    \label{fig: 1csf}
\end{figure*}

\begin{figure*}
    \centering
    \includegraphics[width=\linewidth]{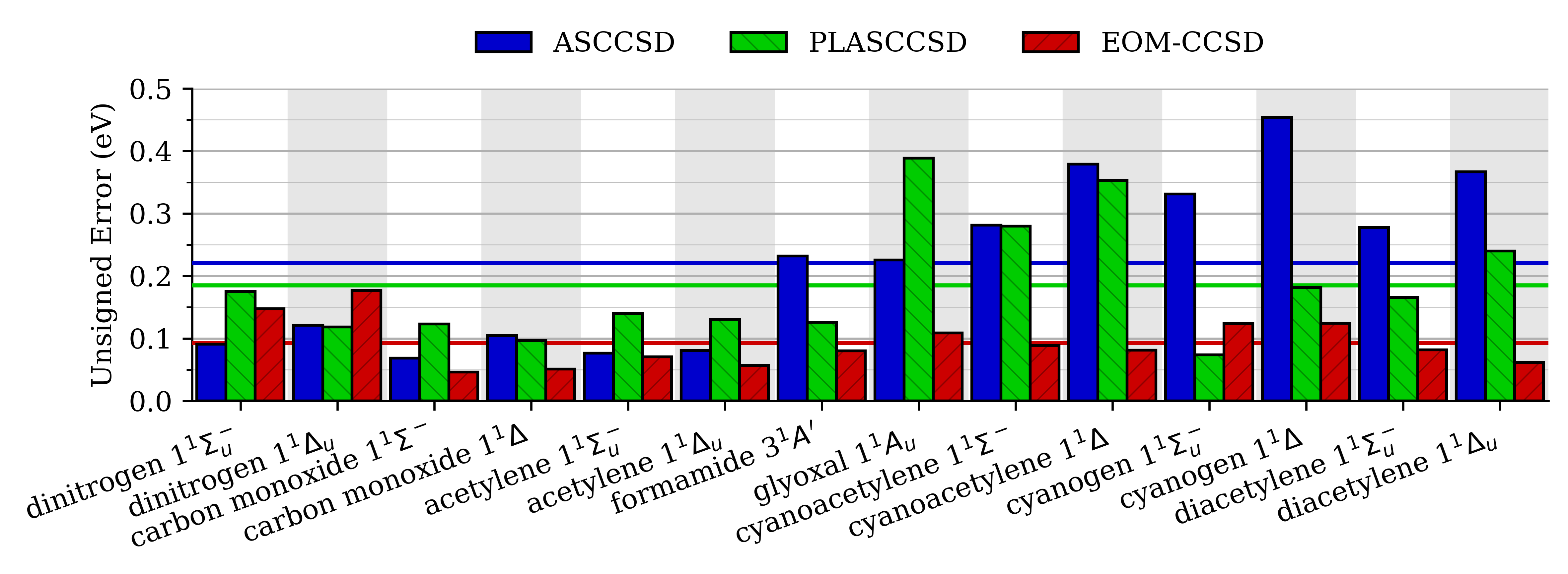}
    \caption{Unsigned errors for ASCCSD, PLASCCSD, and EOM-CCSD in states dominated by two-CSFs. Horizontal lines represent mean unsigned errors for each method.}
    \label{fig: 2csf}
\end{figure*}

Extending on the results from our previous study, we test the accuracy of ASCCSD as compared to EOM-CCSD on the QUEST small and medium molecule valence \textcolor{black}{and Rydberg} excited state benchmarks.\cite{loos2018mountaineering,loos2020mountaineering} Of the 188 singlet excited states in this benchmark, ESMF failed to find a sufficient representation of 10 states after reasonable effort either due to significant root mixing, significant Aufbau contamination, or prominent doubly excited character.  Of the remaining 178 states, ESMF characterized 130 states as single-CSF, 44 as two-CSF, and 4 as three-CSF, the latter of which our ASCC code is currently not set up for. Furthermore, for the $1^1B_g$ state of glyoxal, PLASCC was unable to converge to a physically sensible solution, and closer inspection via (12e,8o) CASSCF revealed the presence of a doubly excited component above our CSF inclusion threshold.  We have therefore omitted this state from our analysis. Finally, due to the computational considerations mentioned above, we limit our investigation of two-CSF states to those containing four or fewer non-hydrogen atoms, thereby reducing the number of two-CSF states to 14. The results for the single- and two-CSF calculations are shown in Figures \ref{fig: 1csf} and \ref{fig: 2csf} respectively \textcolor{black}{and summarized in Table \ref{tab: summary}}, though the interested reader can find \textcolor{black}{additional information} in the Supporting Information.

\begin{figure}
    \centering
    \includegraphics[width=\linewidth]{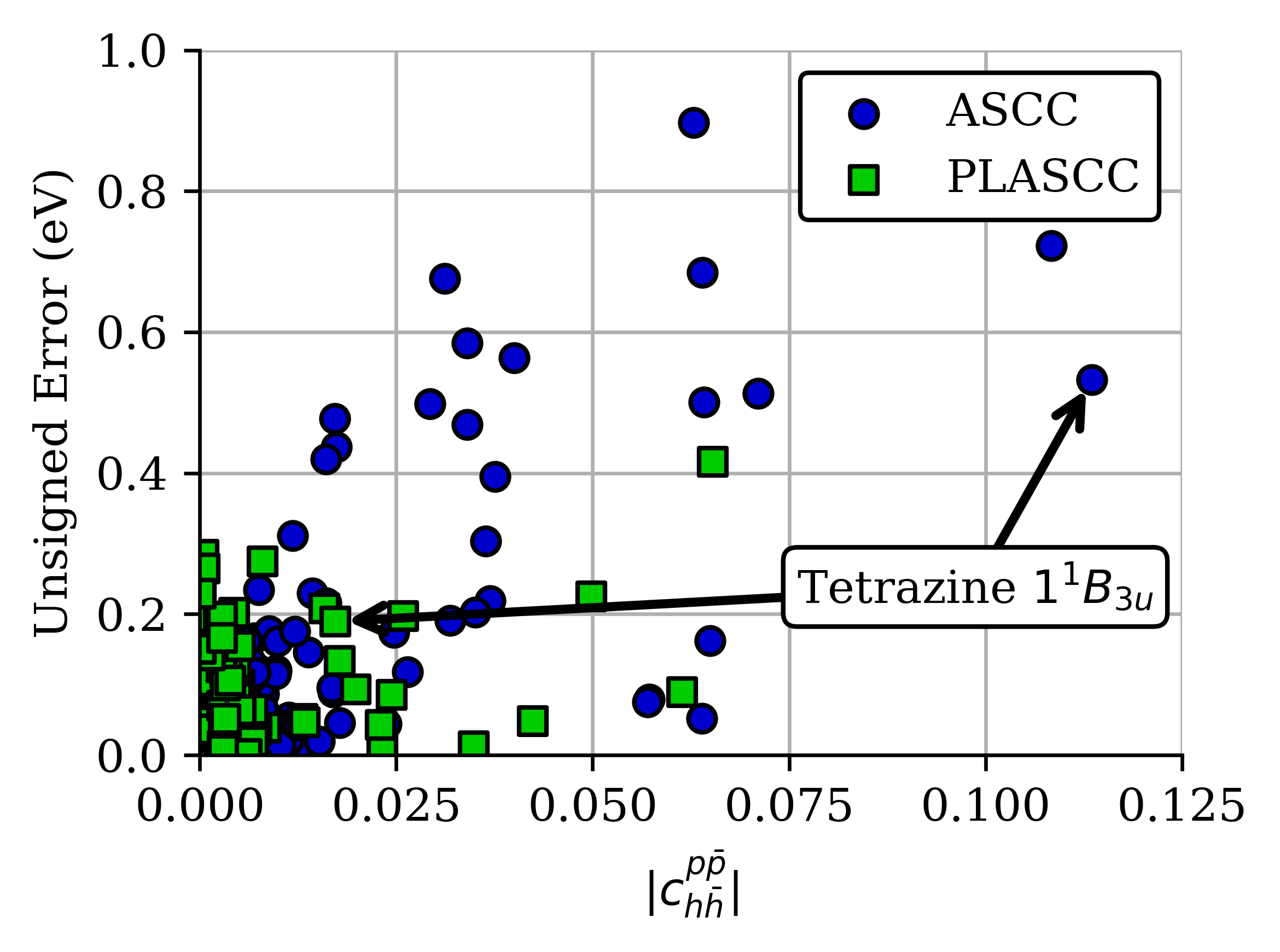}
    \caption{Unsigned errors in excitation energies for ASCCSD and PLASCCSD as a
    function of the unsigned coefficient on the all-primary doubly excited
    determinant in the expanded CC wave function for single-CSF states
    in which this determinant violates symmetry and would be entirely
    absent in FCI.}
    \label{fig: sym}
\end{figure}

\subsubsection{Single-CSF Results}

Interestingly, upon examination of Figure \ref{fig: 1csf}, one finds that the ASCCSD error distribution appears almost bimodal, with a second, small peak centered around an uncharacteristically high $\sim$0.5 eV error, which is in stark contrast to PLASCCSD and EOM-CCSD which both have a single peak centered near zero. With the exception of the states in the second, high error peak, ASCC and PLASCC generally appear to have slightly better centered errors than EOM-CCSD, which tends to err high. PLASCC and EOM-CCSD overall appear to have comparable unsigned errors. 

Closer examination of the states comprising the high error peak in ASCCSD reveals that the vast majority belong to the family of aromatic, six-membered ring molecules. One potential explanation for this observed trend is that these molecules in general tend to have significantly smaller HOMO-LUMO gaps as a result of their extended $\pi$ systems and aromaticity.
\textcolor{black}{While the gap isn't so small that perturbation theory no longer applies, the perturbative portion of the theory is nonetheless larger than in cases with larger gaps and thus}
the cluster amplitudes tend to be larger
and the concerning nonlinear contributions described in
Section \ref{sec: linearization} become more significant.
Some of these contributions are expected to violate point group symmetry,
and, sure enough, we see in Figure \ref{fig: sym} that ASCCSD
has more and more significant violations as measured by the size of the
all-primary double in the wave function expansions as compared to PLASCCSD.
We also see that the worst errors in ASCCSD tend to be accompanied by the
most significant symmetry violations, and that PLASCCSD generally improves both.
In the $1^1B_{3u}$ state of tetrazine, this improvement is particularly
striking, as called out within the plot.
Even with these improvements, it is worth noting that, for a handful of
states, PLASCCSD still lags behind EOM-CCSD in accuracy by $\sim 0.1$ eV.
Due to the larger size of the perturbation in these aromatic ring
molecules, we suspect that these errors are primarily due to the
perturbation order imbalance between ASCCSD and CCSD energies,
as discussed in Section \ref{sec: perb}.

Another notable trend is that PLASCCSD often offers improvements over EOM-CCSD in the sulfur-containing molecules. In fact, \textcolor{black}{excluding the thiophene states which are complicated by the aromaticity concerns above}, PLASCC consistently provides more accurate excitation energies than EOM-CCSD on the \textcolor{black}{remaining 16} sulfur-containing molecules, \textcolor{black}{producing a mean unsigned error of 0.04 eV as compared to EOM-CCSD's 0.12 eV.} Most notably, for the $1^1B_1$ and $1^1B_2$ states of cyclopropenethione, both of which are $n$ to $\pi ^*$ transitions, PLASCCSD offers improvements of $\sim0.3$ eV relative to EOM-CCSD. This improvement in accuracy can potentially be explained by noting that sulfur is significantly more polarizable than the hydrogen, carbon, nitrogen, and oxygen comprising the remaining molecules in the set. As a result, 
\textcolor{black}{the orbitals on the sulfur may have a greater response to any change in the electronic dipole caused by the electronic excitation, which in turn would necessitate a more complete orbital relaxation treatment. For example,} for states like the $1^1B_1$ and $1^1B_2$ of cyclopropenethione where the initially localized electron is delocalized across the molecule upon excitation, \textcolor{black}{the dipole potentially changes enough that} the essentially linearized orbital relaxation treatment of EOM-CCSD is no longer sufficient, thereby explaining EOM-CCSD's unusually large errors on these states.
While EOM-CCSD's orbital relaxation treatment may have been sufficient for the the less polarizable molecules, it is possible that a more robust orbital relaxation treatment is necessary for maintaining accuracy as the polarizability is increased.
Though this preliminary set of data is too limited to draw any decisive conclusions, it will be interesting to examine the effects of polarizability on the accuracy of these methods in the future.

\subsubsection{Two-CSF Results}

Figure \ref{fig: 2csf} makes clear that the augmentations to the 2-CSF approach described in Section \ref{sec: multicsf} significantly improve accuracy relative to our previous work, where errors averaged over 1 eV. \cite{tuckman2024aufbau} Nevertheless, ASCC and PLASCC still significantly underperform relative to EOM-CCSD. Furthermore, while the error of EOM-CCSD remains relatively consistent across the single- and two-CSF states, ASCC and PLASCC appear to have greater average errors in the two-CSF regime as compared to the single-CSF regime. Though at the moment the source of this additional error is not entirely clear, we can offer a few hypotheses.

First, it is altogether possible that the ESMF reference wave function
harbors some of the blame for the decrease in accuracy.
In some of the doubly conjugated molecules, such as glyoxal, doubly
excited components become more prominent, and ESMF lacks these
entirely.
In the future, it would be interesting to examine whether an improved
reference could aide ASCC's performance in multi-CSF states,
especially in cases where a modest amount of double excitation
character is present.

Another possibility again relates to the imbalance between the perturbative orders of the ground and excited state energies. As was mentioned in Section \ref{sec: multicsf}, when transitioning from 1-CSF to 2-CSF the number of missing amplitudes necessary for a complete third order energy significantly increases. For example, when considering the necessary triples amplitudes alone, going from 1-CSF to 2-CSF effectively doubles the number of missing amplitudes, as the excited state reference contains twice as many CSFs. If the size of the energetic correction from these terms scales roughly linearly with the number of missing amplitudes, this then is a potential plausible explanation for the apparent doubling of the average error in the two-CSF relative to the single-CSF results. It will be interesting in the future to examine approaches that move ASCC towards a correct third order energy.

Finally, for PLASCCSD in particular it is possible that the linearization scheme misses some two-CSF-specific detail that is not present in the single-CSF case. The linearization scheme proposed for PLASCC was entirely motivated by single-CSF examples, and it is quite possible that in the two-CSF case this approach either removes a term that shouldn't have been removed, or misses a contribution that would have been wise to remove. A more in depth perturbative analysis of the ASCC ansatz may yield additional insights on the optimal linearization schemes for both the 1- and 2-CSF approaches. 

\begin{figure*}
    \centering
    \includegraphics[width=\linewidth]{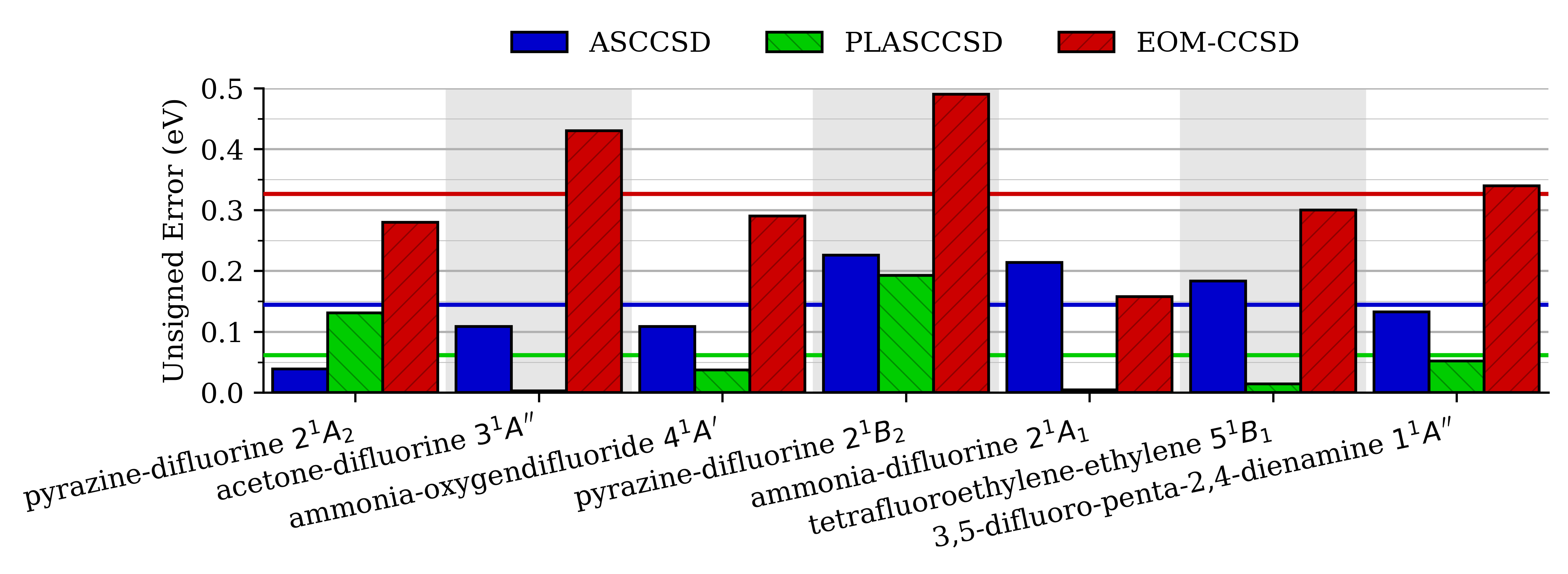}
    \caption{Unsigned errors for ASCCSD, PLASCCSD, and EOM-CCSD in charge transfer excitations. Horizontal lines represent mean unsigned errors for each method.}
    \label{fig: ct}
\end{figure*}

\textcolor{black}{
\setlength{\tabcolsep}{9pt}
\begin{table*}
    \centering
    \textcolor{black}{
    \begin{tabular}{l l c c c}
         Category &             & {ASCCSD} & {PLASCCSD} & {EOM-CCSD} \\ \hline
            Valence and Rydberg & MSE$^{a}$ & \phantom{-}0.12$\pm$0.20 & \phantom{-}0.04$\pm$0.10 & \phantom{-}0.10$\pm$0.09 \\                             
            1-CSF States        & MUE$^{b}$ & \phantom{-}0.15$\pm$0.18 & \phantom{-}0.08$\pm$0.07 & \phantom{-}0.11$\pm$0.08 \\                             
                                & MAE$^{c}$ & \phantom{-}0.90 & \phantom{-}0.42 & \phantom{-}0.50 \\\hline
            Valence and Rydberg & MSE$^{a}$ & \phantom{-}0.22$\pm$0.13 & \phantom{-}0.11$\pm$0.18 & \phantom{-}0.10$\pm$0.04 \\                             
            2-CSF States        & MUE$^{b}$ & \phantom{-}0.22$\pm$0.13 & \phantom{-}0.19$\pm$0.10 & \phantom{-}0.10$\pm$0.04 \\                             
                                & MAE$^{c}$ & \phantom{-}0.45 & \phantom{-}0.39 & \phantom{-}0.18 \\\hline
            Charge Transfer & MSE$^{a}$ & -0.04$\pm$0.16 & \phantom{-}0.05$\pm$0.08 & \phantom{-}0.33$\pm$0.11 \\                             
            1-CSF States        & MUE$^{b}$ & \phantom{-}0.14$\pm$0.07 & \phantom{-}0.06$\pm$0.07 & \phantom{-}0.33$\pm$0.11 \\                             
                                & MAE$^{c}$ & \phantom{-}0.23 & \phantom{-}0.19 & \phantom{-}0.49 \\\hline
    \end{tabular}
    \caption{\textcolor{black}{Summary Statistics in eV Across Different Excitation Categories.}}
    \label{tab: summary}
    \raggedright
    $^{a}$Mean signed error with error bars representing one standard deviation.
    $^{b}$Mean unsigned error with error bars representing one standard deviation.
    $^{c}$Maximum absolute error.
    }
\end{table*}
}

\subsection{Charge Transfer States}

The similar performance of ASCCSD, PLASCCSD, and EOM-CCSD in the 1-CSF valence states is perhaps unsurprising, as state-specificity and better orbital relaxation are not expected to be especially advantageous in such states. However, we do expect these methods to differentiate themselves when these components play a more prominent role, such as in charge transfer excitations. To this end, we compare ASCCSD, PLASCCSD, and EOM-CCSD in a collection of intermolecular charge transfer states for which good reference results are obtainable, many of them from the coupled cluster charge transfer benchmark of Kozma and coworkers. \cite{kozma2020new} The results are \textcolor{black}{shown} in Figure \ref{fig: ct} \textcolor{black}{and summarized in Table \ref{tab: summary}}.

We see that both ASCCSD and PLASCCSD make significant improvements over EOM-CCSD in every state, with the exception of ASCCSD on the ammonia-difluorine charge transfer. These improvements presumably stem from ASCC's robust treatment of orbital relaxation, both in the ESMF reference and through its state-specific $\hat{T}_1$ operator. In contrast, EOM-CCSD is limited to linearized relaxations of the ground state orbitals via the doubles in its response. Interestingly, the partial linearization employed in PLASCCSD continues to improve accuracy, ultimately yielding a MUE below $0.1$ eV. Overall, in contrast to EOM-CCSD which always errs high, ASCCSD and PLASCCSD appear to err high and low in roughly equal amounts. ASCCSD improves over EOM-CCSD's MUE by $\sim$$0.15$ eV, while for PLASCCSD the improvement is $\sim0.25$ eV. Finally, we find it noteworthy that both ASCCSD and PLASCCSD have average errors in charge transfer states that are as good or better than their errors in 1-CSF valence states, suggesting that orbital relaxation challenges have been fully overcome and that higher order terms in the MBPT are likely responsible for the small errors that remain.

\section{Conclusion}

Through a full perturbative analysis of the first order amplitudes and
a preliminary perturbative analysis of the energy, this study has
developed improvements to ASCC that dramatically reduce its errors
in 2-CSF states, bring it on par with EOM-CCSD in 1-CSF states,
and lead it to outperform EOM-CCSD in charge transfer states.
Key improvements include the recognition that small slices of
highly excited amplitudes belong in ASCC's first order
cluster operator, that a partial linearization of the theory
can counteract unwanted side effects of Aufbau suppression,
and that there exist two subtly different ways to construct
the Aufbau-suppressed ansatz.
Through spin-adaptation and improved automated algebra,
better computational performance has expanded the scope
of molecules that can be treated and more firmly established
the utility of versions of ASCC in which the extra terms beyond
those of CCSD all scale as $O(N^5)$ or less.
Moreover, this benchmarking on a greatly expanded set of
molecular excited states has lead to much firmer
conclusions about the theory's accuracy in valence \textcolor{black}{and Rydberg} states and
its advantages in charge transfer states.

Looking forward, there are many exciting avenues for improving and
applying ASCC theory.
First, with the analysis in this study revealing a mismatch in the
perturbative completeness of the ASCC energy as compared to that of
ground state CC, efforts to close this gap are called for.
Although these could take the form of adding more amplitudes and
more iterative $O(N^6)$ terms, non-iterative corrections should
also be investigated.
It is possible that these could close the energetic completeness gap
without altering the iterative part of the algorithm, which would
be especially desirable in multi-CSF ASCC where the number of
additional amplitudes grows with the number of CSFs in the reference.
Further, non-iterative terms that make the ASCC energy complete
to third order would be a stepping stone towards
$O(N^7)$ methods analogous to CCSD(T).

Second, it will be interesting to explore the effectiveness of alternatives
to ESMF as the ASCC reference.
Although the state-specific orbital relaxations of ESMF dovetail nicely with
the state-specificity of ASCC, recent work has suggested that
excited-state-specific CC methods are not particularly sensitive
to orbital relaxations in the reference method, \cite{damour2024state}
presumably thanks to their state-specific $\hat{T}_1$ operators.
Furthermore, by employing more robust reference methods, the error resulting
from the reference versus that from ASCC itself can be more
effectively disentangled.
In particular, it would be interesting to investigate to what extent
ASCC is able to update its own reference by extracting the largest
pieces of the converged ASCC wave function.

Third, ASCC's strong connection to single-reference ground state CC,
along with its state-specificity and size intensivity,
should allow it to interface with local correlation methods.
The corresponding reductions in cost would be especially helpful
in realistic charge transfer applications, in which the donor, bridge,
and acceptor moieties can add up to substantial system sizes,
and where the inclusion of explicit solvation shells may matter.
Beyond charge transfer, it will also be interesting to explore
whether the Aufbau suppression approach can be usefully extended
to core excitations, double excitations, and highly polarizable molecules.

\section{Acknowledgements}

This work was supported by the National Science Foundation,
Award Number 2320936.
Calculations were performed 
using the Savio computational cluster resource provided by the Berkeley Research Computing program at the University of California, Berkeley and the Lawrencium computational cluster resource provided by the IT Division at the Lawrence Berkeley National Laboratory.
H.T. acknowledges that this 
material is based upon work supported by the National Science Foundation 
Graduate Research Fellowship Program under Grant No.\ 
DGE 2146752. Any opinions, findings, and conclusions or recommendations 
expressed in this material are those of the authors and do not necessarily 
reflect the views of the National Science Foundation.

\section{References}
\bibliographystyle{achemso}
\bibliography{main}

\clearpage
\onecolumngrid

\section{Supporting Information}
\renewcommand{\thesection}{S\arabic{section}}
\renewcommand{\theequation}{S\arabic{equation}}
\renewcommand{\thefigure}{S\arabic{figure}}
\renewcommand{\thetable}{S\arabic{table}}
\setcounter{section}{0}
\setcounter{figure}{0}
\setcounter{equation}{0}
\setcounter{table}{0}
\section{Spin Adaptation}

The most recent implementation of ASCC employs spin adaptation in order to significantly reduce both computational cost and the amount of required computer memory. Through the use of the usual Clebsh-Gordon coefficients, one can determine the symmetry allowed CSFs which in turn identify the symmetry allowed cluster amplitudes. When this is paired with the usual antisymmetry of the cluster operators, relationships amongst these amplitudes can be established in order to identify a minimal set of unique cluster amplitudes. For singlet states, this ultimately results in only amplitudes containing as close to equal number of alpha and beta indices as possible being unique and necessary for storage (i.e. $t_{i_\alpha}^{a_{\alpha}}$, $t_{i_\alpha j_\beta}^{a_\alpha b_\beta}$, $t_{i_\alpha j_\beta k_\beta} ^{a_\alpha b_\beta c_\beta}$, etc.). Furthermore, any two amplitudes related by a complete spin flip are exactly equivalent to one another (i.e. $t_{i_\alpha}^{a_{\alpha}}=t_{i_\beta}^{a_{\beta}}$, etc.). Finally, the remaining symmetry allowed amplitudes may be expressed according to the following relationships (where amplitudes on the right side are implied to belong to the set of unique amplitudes for brevity):

\begin{align}
    t_{i_\alpha j_\alpha}^{a_\alpha b_\alpha}&=t_{ij}^{ab}-t_{ji}^{ab}\\
    t_{i_\alpha j_\alpha k_\alpha}^{a_\alpha b_\alpha c_\alpha}&=t_{kij}^{cab}-t_{kij}^{bac}+t_{kij}^{abc}\\
    t_{i_\alpha j_\alpha k_\alpha l_\alpha}^{a_\alpha b_\alpha c_\alpha d_\alpha}&=t_{iljk}^{cdab}-t_{iljk}^{bdac}+t_{iljk}^{bcad}+t_{iljk}^{adbc}-t_{iljk}^{acbd}+t_{iljk}^{abcd}\\
    t_{i_\beta j_\alpha k_\alpha l_\alpha}^{a_\beta b_\alpha c_\alpha d_\alpha}&=\frac{1}{2}\left[t_{ijkl}^{adbc}-t_{ijkl}^{acbd}+t_{ijkl}^{abcd}-t_{ikjl}^{adbc}+t_{ikjl}^{acbd}-t_{ikjl}^{abcd}-t_{iljk}^{cdab}+t_{iljk}^{bdac}-t_{iljk}^{bcad}\right]\\
    t_{i_\alpha j_\alpha k_\alpha l_\alpha m_\alpha}^{a_\alpha b_\alpha c_\alpha d_\alpha e_\alpha}&=t_{lmijk}^{deabc}-t_{lmijk}^{ceabd}+t_{lmijk}^{cdabe}+t_{lmijk}^{beacd}-t_{lmijk}^{bdace}+t_{lmijk}^{bcade}\notag \\ &\quad \quad \quad-t_{lmijk}^{aebcd}+t_{lmijk}^{adbce}-t_{lmijk}^{acbde}+t_{lmijk}^{abcde}\\
    t_{i_\beta j_\alpha k_\alpha l_\alpha m_\alpha}^{a_\beta b_\alpha c_\alpha d_\alpha e_\alpha}&=\frac{1}{2}\left[t_{iljkm}^{adbce}-t_{iljkm}^{aebcd}-t_{iljkm}^{acbde}+t_{iljkm}^{abcde}+t_{imjkl}^{aebcd}-t_{imjkl}^{adbce}+t_{imjkl}^{acbde}-t_{imjkl}^{abcde}\right. \notag \\
    &\quad \quad \quad \left. +t_{lmjki}^{aebcd}-t_{lmjki}^{adbce}+t_{lmjki}^{acbde}-t_{lmjki}^{abcde}\right]
\end{align}

\noindent We note that due to the extensive antisymmetry of especially the higher order cluster operators, these relationships are not unique but nevertheless achieve a correct, singlet spin eigenstate.

\newpage

\section{Automated Code Generation}

Due to the copious amount of terms produced by subdividing the occupied and virtual spaces into primary and nonprimary spaces, ASCC implementations rely heavily on automated code generation. The automated code generation occurs in four distinct phases: term generation, term evaluation, term factorization, and code writing. In the term generation phase, a set of user defined inputs are used to generate a list of terms consisting of symbolic second quantization algebras that require evaluation. During this phase, the user specifies the subdivision of the occupied and virtual spaces (in ASCC's case, the primary versus nonprimary spaces) and defines a list of cluster operators to be included in the equations. The code generator then recursively generates the list of terms necessary for the evaluation of the energy and the residuals corresponding to the included cluster operators. This list of terms is then passed to the second phase of the code generation, the term evaluation.

The term evaluation phase utilizes the ideas introduced in Wick's theorem and the CC diagramatic techniques to evaluate the list of symbolic second quantized algebras and convert them to a list of mathematical tensor contractions. CC diagrams are constructed and connected from bottom to top in a combinatorial manner to generate all possible unique connections. Importantly, however, the code generator recognizes the antisymmetry of the input tensors to avoid redundant contractions that might otherwise be encountered through a true combinatorial expansion of Wick's theorem, just as the diagrammatic CC techniques avoid these terms. Finally, after all possible diagrams are generated, the code generator analyzes the resulting graphs to determine the number of closed loops, hole lines, equivalent lines, equivalent vertices, permutational symmetries, the connectedness of the entire graph, and other important properties in order to determine the sign, coefficient, and permutation on each of the resulting diagrams. After all terms are evaluated, the list of tensor contractions is passed to the third phase of the code generation, the term factorization.

The term factorization phase takes inspiration from the work of Kallay and coworkers, essentially adapting it for use with ASCC's subdivision of primary and nonprimary spaces. Therefore, for more robust details we refer the reader to their work, but we nevertheless will briefly summarize the techniques here. The factorization phase begins by taking each individual term and determining the contraction order which produces the theoretical lowest scaling. Note that by determining the contraction order on a term by term basis rather than by considering all terms at once, more global factorizations such as the $\tau _{ij}^{ab}=t_{ij}^{ab}+t_i^at_j^b$ intermediate commonly found in hand-factorized CC code are missed. Nevertheless, the term by term evaluation still guarantees the correct asymptotic scaling. Each term is then assigned a unique string which contains information regarding the number of internal and external lines of each index type (i.e. primary versus nonprimary, occupied versus virtual, etc.) in each contraction. Then, for each of the different categories of projection (i.e. energy, singles, doubles, etc.), the terms are sorted by this string in order to place terms which are good candidates for forming an intermediate directly next to each other. On a projection by projection basis, all intermediates are then formed by examining the strings of adjacent terms in the sorted term list. Finally, intermediates are compared across projections in order to identify equivalent intermediates across different projection types. With the contraction order and factorization determined, all that remains is the code writing phase of the code generation.

The code writing phase takes the list of intermediates produced in the previous phase and outputs computer code which numerically evaluates those intermediates. In the current implementation, the logic was devised to fit the syntax of NumPy's tensordot function. The code generator utilizes fully antisymmetric tensors and determines how to slice and transpose tensors in order to evaluate the tensor contractions in each intermediate with tensordot calls. The code generator then utilizes some basic logic in order to determine a route through the intermediates that allows for the deletion of an intermediate as quickly as possible in order to save computer memory. Finally, the code generator incorporates the spin adaptation definitions above in order to provide definitions for the non-unique cluster amplitudes. The result of this phase is some code which initializes the different amplitude tensors, calculates the originally requested energy and residual equations, and for convenience, provides the updates for each of the amplitudes based on the diagonal Jacobian approximation.

\newpage

\renewcommand{\arraystretch}{0.75}

\section{Single-CSF States}

\setlength{\tabcolsep}{2pt}

\begin{longtable}{l l |*{1}{S[table-format=2.2]}|*{3}{S[table-format=2.2]}|*{3}{S[table-format=2.2]}|*{2}{S[table-format=2.2]}}
\caption{Single-CSF$^a$ Vertical Excitation Energies (eV)}\\
 & & & \multicolumn{3}{c|}{ASCC} & \multicolumn{3}{c|}{PLASCC} & {EOM-} &  \\
Molecule & State   & {ESMF}         & {One} & {Two} & {Avg.} & {One} & {Two} & {Avg.} & {CCSD} & {Ref.$^{b}$} \\ \hline
\endfirsthead
\caption{Single-CSF$^a$ Vertical Excitation Energies (eV) -- continued from previous page}                                      \\
 & & & \multicolumn{3}{c|}{ASCC} & \multicolumn{3}{c|}{PLASCC} & {EOM-} &  \\
Molecule & State   & {ESMF}         & {One} & {Two} & {Avg.} & {One} & {Two} & {Avg.} & {CCSD} & {Ref.$^{b}$} \\ \hline
\endhead

water                & $1^1B_1$        & 6.46 & 7.50 & 7.50 & 7.50 & 7.51 & 7.51 & 7.51 & 7.45 & 7.53\\
                     & $1^1A_2$        & 8.13 & 9.27 & 9.27 & 9.27 & 9.28 & 9.28 & 9.28 & 9.21 & 9.32\\
                     & $2^1A_1$        & 8.87 & 9.86 & 9.94 & 9.90 & 9.90 & 9.92 & 9.91 & 9.86 & 9.94\\
hydrogen sulfide     & $1^1B_1$        & 5.75 & 6.12 & 6.12 & 6.12 & 6.11 & 6.11 & 6.11 & 6.13 & 6.10\\
                     & $1^1A_2$        & 6.11 & 6.28 & 6.28 & 6.28 & 6.28 & 6.28 & 6.28 & 6.34 & 6.29\\
ammonia              & $1^1A_2$        & 5.56 & 6.42 & 6.47 & 6.45 & 6.46 & 6.47 & 6.46 & 6.46 & 6.48\\
                     & $1^1E$          & 6.99 & 8.03 & 8.03 & 8.03 & 8.06 & 8.06 & 8.06 & 8.03 & 8.08\\
                     & $2^1A_1$        & 8.62 & 9.65 & 9.42 & 9.53 & 9.64 & 9.57 & 9.61 & 9.65 & 9.68\\
                     & $2^1A_2$        & 9.48 & 10.45 & 10.21 & 10.33 & 10.40 & 10.39 & 10.39 & 10.38 & 10.41\\
hydrogen chloride    & $1^1\Pi$        & 7.44 & 7.82 & 7.82 & 7.82 & 7.82 & 7.82 & 7.82 & 7.86 & 7.82\\
dinitrogen           & $1^1\Pi_g$      & 9.79 & 9.64 & 9.64 & 9.64 & 9.46 & 9.46 & 9.46 & 9.49 & 9.41\\
carbon monoxide      & $1^1\Pi$        & 8.48 & 8.66 & 8.66 & 8.66 & 8.59 & 8.59 & 8.59 & 8.67 & 8.57\\
                     & $2^1\Sigma^+$   & 10.95& 11.19 & 10.78 & 10.99 & {---} & 10.88 & 10.88 & 11.17 & 10.94\\
                     & $3^1\Sigma^+$   & 11.26& 11.48 & {---} & 11.48 & 11.51 & 11.65 & 11.58 & 11.71 & 11.52\\
                     & $21^1\Pi$       & 11.72& 11.87 & 11.87 & 11.87 & 11.89 & 11.89 & 11.89 & 11.97 & 11.76\\
ethylene             & $1^1B_{3u}$     & 6.06 & 7.22 & 7.22 & 7.22 & 7.31 & 7.31 & 7.31 & 7.33 & 7.31\\
                     & $1^1B_{1u}$     & 7.24 & 7.88 & 7.88 & 7.88 & 7.88 & 7.88 & 7.88 & 8.04 & 7.93\\
                     & $1^1B_{1g}$     & 6.70 & 7.91 & 7.91 & 7.91 & 8.00 & 8.00 & 8.00 & 8.01 & 8.00\\
formaldehyde         & $1^1A_2$        & 3.09 & 3.95 & 3.95 & 3.95 & 3.95 & 3.95 & 3.95 & 4.02 & 3.99\\
                     & $1^1B_2$        & 6.26 & 7.11 & 7.11 & 7.11 & 7.09 & 7.09 & 7.09 & 7.04 & 7.11\\
                     & $2^1B_2$        & 7.30 & 8.08 & 8.08 & 8.08 & 8.05 & 8.05 & 8.05 & 7.99 & 8.04\\
                     & $2^1A_2$        & 8.00 & 8.72 & 8.72 & 8.72 & 8.69 & 8.69 & 8.69 & 8.61 & 8.65\\
                     & $1^1B_1$        & 8.35 & 9.27 & 9.27 & 9.27 & 9.27 & 9.27 & 9.27 & 9.37 & 9.29\\
thioformaldehyde     & $1^1A_2$        & 1.54 & 2.16 & 2.16 & 2.16 & 2.23 & 2.23 & 2.23 & 2.32 & 2.26$^c$\\
                     & $1^1B_2$        & 5.29 & 5.85 & 5.85 & 5.85 & 5.83 & 5.83 & 5.83 & 5.84 & 5.83\\
                     & $2^1A_1$        & 6.30 & 6.61 & 6.74 & 6.67 & 6.74 & 6.50 & 6.62 & 6.75 & 6.51\\
methanimine          & $1^1A''$        & 4.53 & 5.22 & 5.22 & 5.22 & 5.23 & 5.23 & 5.23 & 5.31 & 5.25\\
acetaldehyde         & $1^1A''$        & 4.38 & 4.30 & 4.30 & 4.30 & 4.28 & 4.28 & 4.28 & 4.36 & 4.34\\
cyclopropene         & $1^1B_1$        & 6.37 & 6.77 & 6.77 & 6.77 & 6.74 & 6.74 & 6.74 & 6.78 & 6.71$^d$\\
                     & $1^1B_2$        & 6.48 & 6.86 & 6.86 & 6.86 & 6.78 & 6.78 & 6.78 & 6.88 & 6.82\\
diazomethane         & $1^1A_2$        & 2.11 & 2.97 & 2.97 & 2.97 & 3.06 & 3.06 & 3.06 & 3.23 & 3.09\\
                     & $1^1B_1$        & 4.87 & 5.31 & 5.31 & 5.31 & 5.34 & 5.34 & 5.34 & 5.43 & 5.35\\
                     & $2^1A_1$        & 5.71 & 5.84 & 5.90 & 5.87 & 5.81 & 5.56 & 5.68 & 5.90 & 5.79\\
formamide            & $1^1A''$        & 4.69 & 5.62 & 5.62 & 5.62 & 5.60 & 5.60 & 5.60 & 5.71 & 5.70\\
                     & $2^1A'$         & 5.87 & 6.73 & 6.84 & 6.78 & 6.78 & 6.76 & 6.77 & 6.83 & 6.67\\
                     & $4^1A'$         & 6.45 & 7.40 & 7.36 & 7.38 & 7.37 & 7.36 & 7.36 & 7.41 & 7.29\\
ketene               & $1^1A_2$        & 3.38 & 3.84 & 3.84 & 3.84 & 3.82 & 3.82 & 3.82 & 3.97 & 3.84\\
                     & $1^1B_1$        & 5.55 & 5.93 & 5.93 & 5.93 & 5.92 & 5.92 & 5.92 & 5.94 & 5.88\\
                     & $2^1A_2$        & 6.52 & 7.10 & 7.10 & 7.10 & {---} & 7.13$^e$ & 7.13 & 7.15 & 7.08\\
nitrosomethane       & $1^1A''$        & 2.35 & 2.04 & 2.04 & 2.04 & 2.02 & 2.02 & 2.02 & 2.00 & 1.99\\
streptocyanine-C1    & $1^1B_2$        & 7.26 & 7.19 & 7.19 & 7.19 & 7.12 & 7.12 & 7.12 & 7.22 & 7.14\\
acetone              & $1^1A_2$        & 3.67 & 4.47 & 4.47 & 4.47 & 4.44 & 4.44 & 4.44 & 4.53 & 4.48\\
                     & $1^1B_2$        & 5.76 & 6.47 & 6.47 & 6.47 & 6.39 & 6.39 & 6.39 & 6.40 & 6.30\\
                     & $2^1A_1$        & 6.60 & 7.52 & 7.49 & 7.51 & 7.42 & 7.53 & 7.47 & 7.46 & 7.36\\
                     & $2^1A_2$        & 6.61 & 7.49 & 7.49 & 7.49 & 7.44 & 7.44 & 7.44 & 7.43 & 7.38\\
                     & $2^1B_2$        & 6.80 & 7.67 & 7.67 & 7.67 & 7.62 & 7.62 & 7.62 & 7.60 & 7.55\\
isobutene            & $1^1B_1$        & 5.43 & 6.41 & 6.41 & 6.41 & 6.44 & 6.44 & 6.44 & 6.44 & 6.39\\
                     & $2^1A_1$        & 6.14 & 7.07 & 7.00 & 7.03 & 7.06 & 6.93 & 6.99 & 7.06 & 7.00\\
thioacetone          & $1^1A_2$        & 1.92 & 2.48 & 2.48 & 2.48 & 2.52 & 2.52 & 2.52 & 2.65 & 2.57\\
                     & $1^1B_2$        & 5.00 & 5.52 & 5.52 & 5.52 & 5.48 & 5.48 & 5.48 & 5.52 & 5.43\\
                     & $2^1A_1$        & 6.22 & 5.94 & 6.18 & 6.06 & 6.04 & 5.88 & 5.96 & 6.07 & 5.98\\
                     & $2^1B_2$        & 5.88 & 6.55 & 6.55 & 6.55 & 6.42 & 6.42 & 6.42 & 6.49 & 6.44\\
                     & $3^1A_1$        & 5.74 & 6.51 & 6.50 & 6.50 & 6.55 & 6.44 & 6.49 & 6.61 & 6.53\\
cyanoformaldehyde    & $1^1A''$        & 2.89 & 3.82 & 3.82 & 3.82 & 3.79 & 3.79 & 3.79 & 3.94 & 3.84\\
                     & $2^1A''$        & 5.36 & 6.56 & 6.56 & 6.56 & 6.49 & 6.49 & 6.49 & 6.77 & 6.54\\
propynal             & $1^1A''$        & 2.94 & 3.80 & 3.80 & 3.80 & 3.76 & 3.76 & 3.76 & 3.93 & 3.82\\
                     & $2^1A''$        & 4.69 & 5.43 & 5.43 & 5.43 & 5.57 & 5.57 & 5.57 & 5.77 & 5.62\\
thiopropynal         & $1^1A''$        & 1.39 & 1.96 & 1.96 & 1.96 & 2.00 & 2.00 & 2.00 & 2.17 & 2.06\\
cyclopropenone       & $1^1B_1$        & 4.14 & 4.40 & 4.40 & 4.40 & 4.27 & 4.27 & 4.27 & 4.47 & 4.23\\
                     & $1^1A_2$        & 5.29 & 5.65 & 5.65 & 5.65 & 5.53 & 5.53 & 5.53 & 5.62 & 5.56\\
                     & $1^1B_2$        & 5.68 & 6.37 & 6.37 & 6.37 & 6.30 & 6.30 & 6.30 & 6.27 & 6.19\\
                     & $3^1B_2$        & 6.40 & 7.02 & 7.02 & 7.02 & 6.92 & 6.92 & 6.92 & 6.96 & 6.86\\
                     & $2^1A_1$        & 6.27 & 7.01 & 7.02 & 7.01 & 6.91 & 6.99 & 6.95 & 6.95 & 6.87\\
                     & $3^1A_1$        & 8.57 & 8.53 & 8.69 & 8.61 & 8.53 & 7.87$^e$ & 8.20 & 8.36 & 8.29\\
cyclopropenethione   & $1^1A_2$        & 3.22 & 3.44 & 3.44 & 3.44 & 3.40 & 3.40 & 3.40 & 3.54 & 3.45\\
                     & $1^1B_1$        & 3.35 & 3.49 & 3.49 & 3.49 & 3.42 & 3.42 & 3.42 & 3.77 & 3.42\\
                     & $1^1B_2$        & 4.33 & 4.59 & 4.59 & 4.59 & 4.59 & 4.59 & 4.59 & 4.95 & 4.64\\
                     & $2^1B_2$        & 4.67 & 5.27 & 5.27 & 5.27 & 5.25 & 5.25 & 5.25 & 5.27 & 5.21\\
                     & $3^1B_2$        & 5.42 & 5.96 & 5.96 & 5.96 & 5.86 & 5.86 & 5.86 & 5.93 & 5.84\\
methylenecyclopropene& $1^1B_2$        & 3.69 & 4.25 & 4.25 & 4.25 & 4.30 & 4.30 & 4.30 & 4.55 & 4.31\\
                     & $1^1B_1$        & 4.34 & 5.35 & 5.35 & 5.35 & 5.41 & 5.41 & 5.41 & 5.37 & 5.35\\
                     & $1^1A_2$        & 4.80 & 5.86 & 5.86 & 5.86 & 5.92 & 5.92 & 5.92 & 5.90 & 5.88\\
                     & $2^1A_1$        & 5.36 & 6.21 & 6.11 & 6.16 & 6.20 & 6.14 & 6.17 & 6.18 & 6.15\\
acrolein             & $1^1A''$        & 3.16 & 3.78 & 3.78 & 3.78 & 3.68 & 3.68 & 3.68 & 3.90 & 3.74\\
                     & $2^1A'$         & 6.57 & 6.93 & 6.91 & 6.92 & 6.74 & 6.81 & 6.77 & 6.86 & 6.70\\
                     & $3^1A'$         & 6.03 & 7.01 & 7.01 & 7.01 & 6.94 & 7.03 & 6.98 & 7.11 & 7.00\\
butadiene            & $1^1B_u$        & 6.01 & 6.35 & 6.35 & 6.35 & 6.28 & 6.28 & 6.28 & 6.37 & 6.27\\
                     & $1^1B_g$        & 5.49 & 6.33 & 6.33 & 6.33 & 6.39 & 6.39 & 6.39 & 6.30 & 6.27\\
                     & $2^1A_g$        & 6.91 & 7.24 & 7.19 & 7.21 & {---} & 6.71 & 6.71 & 7.09 & 6.59\\
                     & $1^1A_u$        & 5.82 & 6.65 & 6.65 & 6.65 & 6.74 & 6.74 & 6.74 & 6.62 & 6.59\\
                     & $2^1A_u$        & 5.89 & 6.80 & 6.80 & 6.80 & 6.87 & 6.87 & 6.87 & 6.78 & 6.74\\
                     & $2^1B_u$        & 7.38 & 8.02$^e$ & 8.01$^e$ & 8.02 & 7.96 & 7.96 & 7.96 & 7.93 & 7.87\\
pyrrole              & $1^1A_2$        & 4.48 & 5.21 & 5.21 & 5.21 & 5.24 & 5.24 & 5.24 & 5.22 & 5.14\\
                     & $1^1B_1$        & 5.60 & 6.00 & 6.00 & 6.00 & 5.99 & 5.99 & 5.99 & 5.91 & 5.87\\
                     & $2^1A_2$        & 5.19 & 5.96 & 5.96 & 5.96 & 6.02 & 6.02 & 6.02 & 5.99 & 5.93\\
                     & $1^1B_2$        & 5.80 & 6.39 & 6.40 & 6.40 & 6.23 & 6.23 & 6.23 & 6.37 & 6.28\\
                     & $2^1B_2$        & 6.74 & 7.05 & 7.05 & 7.05 & 7.04 & 7.04 & 7.04 & 7.08 & 6.83\\
furan                & $1^1A_2$        & 5.26 & 6.08 & 6.08 & 6.08 & 6.12 & 6.12 & 6.12 & 6.07 & 6.00\\
                     & $1^1B_2$        & 6.06 & 6.51 & 6.51 & 6.51 & 6.29$^e$ & 6.31$^e$ & 6.30 & 6.53 & 6.39\\
                     & $1^1B_1$        & 5.68 & 6.62 & 6.62 & 6.62 & 6.68 & 6.68 & 6.68 & 6.61 & 6.56\\
                     & $2^1A_2$        & 5.91 & 6.78 & 6.78 & 6.78 & 6.85 & 6.85 & 6.85 & 6.80 & 6.74\\
                     & $2^1B_2$        & 6.84 & 7.20 & 7.20 & 7.20 & 7.47 & 7.47 & 7.47 & 7.47 & 7.40\\
cyclopentadiene      & $1^1B_2$        & 5.33 & 5.67 & 5.67 & 5.67 & 5.59 & 5.59 & 5.59 & 5.71 & 5.60\\
                     & $1^1A_2$        & 4.90 & 5.78 & 5.78 & 5.78 & 5.83 & 5.83 & 5.83 & 5.74 & 5.70\\
                     & $1^1B_1$        & 5.37 & 6.39 & 6.39 & 6.39 & 6.47 & 6.47 & 6.47 & 6.36 & 6.34\\
                     & $2^1A_2$        & 5.50 & 6.43 & 6.43 & 6.43 & 6.51 & 6.51 & 6.51 & 6.42 & 6.39\\
                     & $2^1B_2$        & 5.64 & 6.37 & 6.37 & 6.37 & 6.65 & 6.65 & 6.65 & 6.58 & 6.55\\
thiophene            & $1^1B_2$        & 5.90 & 6.22 & 6.22 & 6.22 & 6.08 & 6.08 & 6.08 & 6.20 & 6.06\\
                     & $1^1A_2$        & 5.46 & 6.17 & 6.17 & 6.17 & 6.19 & 6.19 & 6.19 & 6.13 & 6.06\\
                     & $1^1B_1$        & 5.76 & 6.40 & 6.40 & 6.40 & 5.75 & 5.75 & 5.75 & 6.31 & 6.17\\
                     & $2^1A_2$        & 6.37 & 6.44 & 6.44 & 6.44 & 6.22 & 6.22 & 6.22 & 6.37 & 6.31\\
                     & $2^1B_1$        & 6.31 & 6.57 & 6.57 & 6.57 & 6.41 & 6.41 & 6.41 & 6.46 & 6.41\\
                     & $2^1B_2$        & 6.96 & 7.56 & 7.56 & 7.56 & 7.59 & 7.59 & 7.59 & 7.52 & 7.44\\
imidazole            & $1^1A''$        & 4.84 & 5.65 & 5.65 & 5.65 & 5.68 & 5.68 & 5.68 & 5.68 & 5.60\\
                     & $2^1A'$         & 6.04 & 6.56 & 6.68 & 6.62 & 5.98 & 6.44 & 6.21 & 6.58 & 6.43\\
                     & $2^1A''$        & 5.53 & 6.40 & 6.40 & 6.40 & 6.49 & 6.49 & 6.49 & 6.47 & 6.42\\
benzene              & $1^1E_{1g}$     & 5.89 & 6.58 & 6.58 & 6.58 & 6.58 & 6.58 & 6.58 & 6.49 & 6.46\\
tetrazine            & $1^1B_{3u}$     & 3.33 & 3.03 & 3.03 & 3.03 & 2.69 & 2.69 & 2.69 & 2.65 & 2.50\\
                     & $1^1A_u$        & 5.31 & 4.38 & 4.38 & 4.38 & 3.97 & 3.97 & 3.97 & 3.93 & 3.70\\
                     & $1^1B_2$u       & 6.02 & 5.97 & 5.97 & 5.97 & 5.48 & 5.48 & 5.48 & 5.40 & 5.25\\
                     & $1^1B_2$g       & 6.64 & 6.40 & 6.40 & 6.40 & 5.54 & 5.54 & 5.54 & 5.88 & 5.50\\
pyridazine           & $1^1B_1$        & 4.30 & 4.16 & 4.16 & 4.16 & 3.92 & 3.92 & 3.92 & 4.03 & 3.86\\
                     & $1^1B_2$        & 6.55 & 6.37 & 6.37 & 6.37 & 6.26 & 6.26 & 6.26 & 6.25 & 6.06\\
                     & $2^1B_1$        & 7.48 & 6.91 & 6.91 & 6.91 & 6.60 & 6.60 & 6.60 & 6.66 & 6.41\\
pyrazine             & $1^1B_{3u}$     & 4.94 & 4.70 & 4.70 & 4.70 & 4.32 & 4.32 & 4.32 & 4.35 & 4.19\\
                     & $1^1A_u$        & 6.53 & 5.54 & 5.54 & 5.54 & 5.18 & 5.18 & 5.18 & 5.19 & 4.98\\
                     & $1^1B_{2u}$     & 5.54 & 5.55 & 5.55 & 5.55 & 5.25 & 5.25 & 5.25 & 5.18 & 5.05\\
                     & $2^1A_g$        & 7.64 & 7.04 & 7.04 & 7.04 & 6.82 & 6.81 & 6.81 & 6.66 & 6.53\\
                     & $1^1B_{1g}$     & 8.67 & 7.43 & 7.43 & 7.43 & 6.95 & 6.95 & 6.95 & 7.10 & 6.75\\
                     & $2^1B_{1g}$     & 6.40 & 7.20 & 7.20 & 7.20 & 7.24 & 7.24 & 7.24 & 7.17 & 7.14\\
                     & $2^1B_{2u}$     & 8.17 & 7.61 & 7.61 & 7.61 & 7.41 & 7.41 & 7.41 & 7.27 & 7.13\\
pyridine             & $1^1B_1$        & 4.73 & 5.10 & 5.10 & 5.10 & 5.01 & 5.01 & 5.01 & 5.19 & 5.00\\
                     & $1^1A_2$        & 5.62 & 5.63 & 5.63 & 5.63 & 5.46 & 5.46 & 5.46 & 5.61 & 5.41\\
                     & $2^1A_2$        & 6.08 & 6.83 & 6.83 & 6.83 & 6.86 & 6.86 & 6.86 & 6.78 & 6.75\\
                     & $2^1B_1$        & 6.52 & 7.41 & 7.41 & 7.41 & 7.46 & 7.46 & 7.46 & 7.34 & 7.32\\
pyrimidine           & $1^1B_1$        & 5.19 & 4.88 & 4.88 & 4.88 & 4.63 & 4.63 & 4.63 & 4.66 & 4.48\\
                     & $1^1A_2$        & 5.88 & 5.35 & 5.35 & 5.35 & 5.03 & 5.03 & 5.03 & 5.06 & 4.88\\
                     & $2^1B_1$        & 7.72 & 6.87 & 6.87 & 6.87 & 6.46 & 6.46 & 6.46 & 6.52 & 6.29\\
                     & $2^1B_2$        & 7.33 & 7.03 & 7.03 & 7.03 & 6.86 & 6.86 & 6.86 & 6.72 & 6.59\\
triazine             & $1^1E'$         & 7.86 & 7.63 & 7.63 & 7.63 & 7.44 & 7.44 & 7.44 & 7.29 & 7.21\\ \hline
MSE$^f$              &                 & -0.35& & & 0.12 & & & 0.04 & 0.10 & \\
Std. Dev.            &                 & 0.66 & & & 0.20 & & & 0.10 & 0.09 & \\
MUE$^g$              &                 & 0.65 & & & 0.15 & & & 0.08 & 0.11 & \\
Std. Dev.            &                 & 0.36 & & & 0.18 & & & 0.07 & 0.08 & \\
\multicolumn{10}{p{0.9\linewidth}}{\raggedright \footnotesize $^a$States where one ESMF singular values is $>$0.2. $^b$The reference is exFCI for molecules with three or fewer non-hydrogen atoms and EOM-CCSDT otherwise unless explicitly stated. $^c$EOM-CCSDTQ reference. $^d$EOM-CCSDT reference. $^e$States are loosely converged, but energetically change below precision reported iteration by iteration. $^f$Mean signed error (MSE). $^g$Mean unsigned error (MUE).}
\end{longtable}

\newpage

\section{Two-CSF States}

\setlength{\tabcolsep}{2pt}

\begin{longtable}{l l |*{1}{S[table-format=2.2]} |*{3}{S[table-format=2.2]}|*{3}{S[table-format=2.2]}|*{2}{S[table-format=2.2]}}
\caption{Two-CSF$^a$ Vertical Excitation Energies (eV)}\\
 & & & \multicolumn{3}{c|}{ASCC} & \multicolumn{3}{c|}{PLASCC} & {EOM-} &  \\
Molecule & State       & {ESMF}     & {One} & {Two} & {Avg.} & {One} & {Two} & {Avg.} & {CCSD} & {Ref.$^{b}$} \\ \hline
\endfirsthead
\caption{Two-CSF$^a$ Vertical Excitation Energies (eV) -- continued from previous page}                                      \\
 & & & \multicolumn{3}{c|}{ASCC} & \multicolumn{3}{c|}{PLASCC} & {EOM-} &  \\
Molecule & State   & {ESMF}         & {One} & {Two} & {Avg.} & {One} & {Two} & {Avg.} & {CCSD} & {Ref.$^{b}$} \\ \hline
\endhead

dinitrogen         & $1^1\Sigma _u^-$  & 8.36 & 10.14 & 10.14 & 10.14 & 10.23 & 10.23 & 10.23 & 10.20 & 10.05\\
                   & $1^1\Delta _u$    & 8.92 & 10.55 & 10.55 & 10.55 & 10.55 & 10.55 & 10.55 & 10.61 & 10.43\\
carbon monoxide    & $1^1\Sigma ^-$    & 8.36 & 10.12 & 10.12 & 10.12 & 10.17 & 10.17 & 10.17 & 10.10 & 10.05\\
                   & $1^1\Delta$       & 8.55 & 10.26 & 10.26 & 10.26 & 10.26 & 10.26 & 10.26 & 10.21 & 10.16\\
acetylene          & $1^1\Sigma _u^-$  & 6.12 & 7.28 & 7.28 & 7.28 & 7.34 & 7.34 & 7.34 & 7.27 & 7.20\\
                   & $1^1\Delta _u$    & 6.42 & 7.59 & 7.59 & 7.59 & 7.64 & 7.64 & 7.64 & 7.57 & 7.51\\
formamide          & $3^1A'$           & 8.02 & 7.84 & 7.90 & 7.87 & 7.70$^c$ & 7.33$^c$ & 7.51 & 7.72 & 7.64\\
glyoxal            & $1^1A_u$          & 3.26 & 3.13 & 3.13 & 3.13 & 2.51 & 2.51 & 2.51 & 3.01 & 2.90\\
cyanoacetylene     & $1^1\Sigma ^-$    & 5.02 & 6.19 & 6.19 & 6.19 & 6.19 & 6.19 & 6.19 & 6.00 & 5.91$^d$\\
                   & $1^1\Delta$       & 5.31 & 6.55 & 6.55 & 6.55 & 6.52 & 6.52 & 6.52 & 6.25 & 6.17$^d$\\
cyanogen           & $1^1\Sigma _u^-$  & 5.43 & 6.84 & 6.84 & 6.84 & 6.58 & 6.58 & 6.58 & 6.63 & 6.51$^d$\\
                   & $1^1\Delta _u$    & 5.78 & 7.22 & 7.22 & 7.22 & 6.95 & 6.95 & 6.95 & 6.89 & 6.77$^d$\\
diacetylene        & $1^1\Sigma _u^-$  & 4.68 & 5.71 & 5.71 & 5.71 & 5.60 & 5.60 & 5.60 & 5.51 & 5.43$^d$\\
                   & $1^1\Delta _u$    & 4.94 & 6.06 & 6.06 & 6.06 & 5.93$^c$ & 5.93$^c$ & 5.93 & 5.75 & 5.69$^d$\\
 
\hline 
MSE$^e$            &                   & -0.95 & & & 0.22 & & & 0.11 & 0.10 & \\
Std. Dev.          &                   & 0.65 & & & 0.13 & & & 0.18 & 0.04 & \\
MUE$^f$            &                   & 1.05 & & & 0.22 & & & 0.19 & 0.10 & \\
Std. Dev.          &                   & 0.44 & & & 0.13 & & & 0.10 & 0.04 & \\
\multicolumn{10}{p{0.9\linewidth}}{\raggedright \footnotesize $^a$States where two ESMF singular values are $>$0.2. $^b$The reference is exFCI for molecules with three or fewer non-hydrogen atoms and EOM-CCSDT otherwise unless explicitly stated. $^c$States are loosely converged, but energetically change below precision reported iteration by iteration. $^d$EOM-CCSDTQ reference. $^e$Mean signed error (MSE). $^f$Mean unsigned error (MUE).}
\end{longtable}

\newpage

\section{Charge Transfer States}

\setlength{\tabcolsep}{2pt}

\begin{longtable}{l l |*{1}{S[table-format=2.2]}|*{3}{S[table-format=2.2]}|*{3}{S[table-format=2.2]}|*{2}{S[table-format=2.2]}}
\caption{Charge Transfer Vertical Excitation Energies (eV)}\\
 & & & \multicolumn{3}{c|}{ASCC} & \multicolumn{3}{c|}{PLASCC} & {EOM-} &  \\
Molecule & State   & {ESMF}         & {One} & {Two} & {Avg.} & {One} & {Two} & {Avg.} & {CCSD} & {Ref.$^{b}$} \\ \hline
\endfirsthead
\caption{Charge Transfer Vertical Excitation Energies (eV) -- continued from previous page}                                      \\
 & & & \multicolumn{3}{c|}{ASCC} & \multicolumn{3}{c|}{PLASCC} & {EOM-} &  \\
Molecule & State   & {ESMF}         & {One} & {Two} & {Avg.} & {One} & {Two} & {Avg.} & {CCSD} & {Ref.$^{b}$} \\ \hline
\endhead

ammonia-difluorine                & $2^1A_1$     & 7.51 & 9.17 & 9.16 & 9.16 & 9.37 & 9.37 & 9.37 & 9.54 & 9.38\\
acetone-difluorine                & $3^1A''$     & 4.17 & 5.74 & 5.74 & 5.74 & 5.85 & 5.85 & 5.85 & 6.28 & 5.85\\
pyrazine-difluorine               & $2^1B_2$     & 6.42 & 6.51 & 6.51 & 6.51 & 6.47 & 6.47 & 6.47 & 6.77 & 6.28\\
                                  & $2^1A_2$     & 4.83 & 6.41 & 6.41 & 6.41 & 6.58 & 6.58 & 6.58 & 6.73 & 6.45\\
ammonia-oxygendifluoride          & $4^1A'$      & 5.41 & 6.95 & 6.92 & 6.93 & 7.01 & 6.99 & 7.00 & 7.33 & 7.04\\
tetrafluoroethylene-ethylene      & $5^1B_1$     & 10.53 & 10.39 & 10.39 & 10.39 & 10.58 & 10.58 & 10.58 & 10.87 & 10.57\\
3,5-difluoro-penta-2,4-dienamine  & $1^1A''$     & 6.44 & 6.84 & 6.84 & 6.84 & 6.76 & 6.76 & 6.76 & 7.05 & 6.71$^b$\\

\hline
MSE$^c$      & & -1.00 & & & -0.04 & & & 0.05 & 0.33 & \\
Std. Dev. & & 0.89 & & & 0.16 & & & 0.08 & 0.11 & \\
MUE$^d$     &  & 1.04 & & & 0.14 & & & 0.06 & 0.33 & \\
Std. Dev. & & 0.84 & & & 0.07 & & & 0.07 & 0.11 & \\
\multicolumn{10}{p{0.9\linewidth}}{\raggedright \footnotesize $^a$The reference is EOM-CCSDT otherwise unless explicitly stated. $^b$LR-CC3 reference. $^c$Mean signed error (MSE). $^d$Mean unsigned error (MUE).}
\end{longtable}

\newpage

\section{Molecular Geometries}

Our ammonia-difluorine and 3,5-difluoro-penta-2,4-dienamine geometries are shown below in Angstroms.  The other geometries can be found in the supporting information of the original QUEST benchmarks and the charge transfer benchmark of Kozma and coworkers.\\

\noindent \bf{ammonia-difluorine}
\begin{verbatim}
N   0.0000000000   0.0000000000   0.1277920000                                                      
H   0.0000000000   0.9318900000  -0.2981820000                                                     
H   0.8070400000  -0.4659450000  -0.2981820000                                                     
H  -0.8070400000  -0.4659450000  -0.2981820000                                                     
F   0.0000000000   0.0000000000   5.1277920000                                                      
F   0.0000000000   0.0000000000   6.5597920000
\end{verbatim}

\noindent \bf{3,5-difluoro-penta-2,4-dienamine}
\begin{verbatim}
N  -2.4131162878  -0.2210598931   0.0000000000            
H  -2.4528357410   0.4050762251   0.8082210202            
H  -2.4528357410   0.4050762251  -0.8082210202            
C  -1.1041315277  -0.8778592602   0.0000000000            
H  -1.0655737822  -1.5422814797   0.8816442239            
H  -1.0655737822  -1.5422814797  -0.8816442239            
C   0.0890809140   0.0602768896   0.0000000000            
H  -0.1008665091   1.1394594998   0.0000000000            
C   1.3828140203  -0.3289756263   0.0000000000            
F   2.3589882632   0.6084301804   0.0000000000            
C   1.8433535010  -1.7101293480   0.0000000000            
H   1.0773275218  -2.4882915857   0.0000000000            
C   3.1174870791  -2.1517822876   0.0000000000            
F   4.1935798301  -1.3571375780   0.0000000000            
H   3.3747047712  -3.2142891283   0.0000000000  
\end{verbatim}

\renewcommand{\thesection}{\Roman{section}}
\setcounter{section}{6}

\end{document}